# Observation of discrete conventional Caroli-de Gennes-Matricon states in the vortex core of single-layer FeSe/$SrTiO_3$

Chen Chen[1,4†], Qin Liu[1,2,4†], Wei-Cheng Bao[3,6†], Yajun Yan[1,4], Qiang-Hua Wang[3,4*], Tong Zhang[1,4*], Donglai Feng[1,4,5*]

[1] State Key Laboratory of Surface Physics, Department of Physics, and Advanced Materials Laboratory, Fudan University, Shanghai 200438, China

[2] Science and Technology on Surface Physics and Chemistry Laboratory, Mianyang, Sichuan 621908, China

[3] National Laboratory of Solid State Microstructures & School of Physics, Nanjing University, Nanjing, 210093, China

[4] Collaborative Innovation Center of Advanced Microstructures, Nanjing 210093, China

[5] Hefei National Laboratory for Physical Science at Microscale and Department of Physics, University of Science and Technology of China, Hefei, Anhui 230026, China

[6] Zhejiang University of Water Resources and Electric Power, Hangzhou 310018, China

[†]These authors contributed equally.

*Corresponding authors. Email: qhwang@nju.edu.cn, tzhang18@fudan.edu.cn, dlfeng@fudan.edu.cn

**Using low-temperature scanning tunneling microscopy (STM), we studied the vortex states of single-layer FeSe film on $SrTiO_3$ (100) substrate, and the local behaviors of superconductivity at sample boundaries. We clearly observed multiple discrete Caroli-de Gennes-Matricon (CdGM) states in the vortex core, and quantitative analysis shows their energies well follow the formula: $E = \mu\Delta^2/E_F$, where $\mu$ is a half integer (±1/2, ±3/2, ±5/2…) and $\Delta$ is the mean superconducting gap over the Fermi surface. Meanwhile, a fully gapped spectrum without states near zero bias is observed at $[110]_{Fe}$ oriented boundary of 1 ML and 2 ML FeSe films, and atomic step edge of 1 ML FeSe. Accompanied with theoretical calculations, our results indicate a *s*-wave pairing without sign-change in the high-$T_C$ FeSe/$SrTiO_3$ superconductor.**

In superconductors, localized quasi-particle states at the boundaries such as magnetic vortices and superconducting/non-superconducting (S/N) interfaces, can provide critical information on electron pairing. As for conventional *s*-wave superconductor, there are so called Caroli-de Gennes-Matricon (CdGM) states in the vortex cores [1] with $E = \mu\Delta^2/E_F$, where $\mu$ is a *half* integer (±1/2, ±3/2…); while for superconductors with sign-changing order parameter, such as *d*-wave and *p*-wave, the vortex states have qualitatively different energy characteristic [2] (e.g. $E$=0 state is expected for chiral *p*-wave superconductor). However, in practice, the identification of CdGM states has long been hampered by their small energy spacing ($\delta E \sim \Delta^2/E_F$), which is in the $\mu$eV range for conventional low-$T_C$ superconductors [3~4]. Progresses are only achieved recently in the iron-based superconductor Fe(Te,Se) [5~7] and (Li,Fe)OHFeSe [8,9], in which discrete vortex states accompanied by a zero-bias conductance peak (ZBCP) were observed. In particular, quantized conductance of the ZBCP has been observed in (Li,Fe)OHFeSe [9], indicating its topological nature and the presence of Majorana zero mode. Meanwhile, zero-energy or dispersive Andreev bound states (ABS) at S/N boundaries are expected for certain sign-change (e.g., *d*-wave [10]) or topological superconductors [11], but not for *s*-wave superconductors. Nonetheless the detection of ABS relies on clean S/N boundaries and well resolved gap spectrum, which are non-trivial in practice.

Single-layer FeSe on $SrTiO_3$ probably has the highest $T_C$ (≥65K) among all the Fe-based superconductors [12~19]. Its pairing symmetry is thus of great importance [20]. Previous STM study has suggested a plain *s*-wave pairing from impurity effect and quasi-particle interference (QPI) [18]. Nonetheless, a recent theoretical work shows that a nodeless *d*-wave pairing is possible if there is band hybridization induced by small spin-orbit coupling (SOC) [21]. Moreover, since calculations did not show topological band structure near the Fermi level of single-layer FeSe/$SrTiO_3$ [22], it thus serves as an important counterpart of the topologically non-trivial Fe(Se,Te) [6,7,23,24] and (Li,Fe)OHFeSe [8]. Examining the vortex states of FeSe/$SrTiO_3$ would provide information on both the pairing symmetry and factors for the presence/absence of Majorana zero mode.

Here we report a low-temperature STM study on the vortex states and S/N boundaries of single-layer FeSe/$SrTiO_3$(100). We observed multiple discrete CdGM states with energies of $E= \mu(\delta E)$, where $\mu$ is a half integer and $\delta E = \Delta^2/E_F$. Quantitatively, $\delta E$ can be accounted by an anisotropic superconducting gap with a mean size of $\Delta_0$, which also explains the double coherence peaks in the tunneling spectrum. Our detailed model calculation shows the observed CdGM states agree with plain *s*-wave but disfavors *d*-wave pairing, which is further supported by observation of full superconducting gap at both $[110]_{Fe}$ oriented 1ML/2ML FeSe boundary and 1ML FeSe step edge. Our results provide critical information on the pairing in this remarkable interfacial superconductor, and suggest the importance of out-of-plane coupling in realizing topological superconductivity in iron-based superconductors.

The FeSe films were grown by co-evaporating selenium (99.999%) and iron (99.995%) (flux ratio ~10:1) on Nb(0.5%) doped $SrTiO_3$(001) substrate at 400°C and post-annealed at 500°C to improve crystallinity. STM measurement was conducted in a cryogenic STM (UNISOKU) with a base temperature of $T$=0.4K and an electron temperature ($T_{elec}$) of 1.18K (see Part-I of Supplementary Materials (SM) [25]). Pt/Ir tips were used after treatment on Au(111) surface. dI/dV spectra were obtained by standard lock-in technique with $f$ = 714Hz.

Fig. 1(a) shows a typical topography of FeSe/$SrTiO_3$ with nominal thickness of 1.3 ML, and Fig. 1(b) is a zoomed image of 1ML FeSe region. The tunneling spectra taken on defect-free area displays a full superconducting gap with flat bottom and two pairs of coherence peaks (Fig. 1(c)). Under a vertical magnetic field of B=10T, vortices show up in the zero-bias dI/dV mapping (Fig. 1(d)). Some of the vortices are pinned by surface defects which can induce strong in-gap state (Fig. S3), as indicated by arrows in Figs. 1(b) and 1(d); while the ones marked by circles display clean local superconducting gap at zero field and we refer them as "free" vortices (see Part-II of SM for detailed gap measurement of free vortices regions). The superconducting coherence length ($\xi$) extracted from vortex mapping is 2.03~2.45nm (Fig. S4), which is significantly smaller than averaged inter-vortex spacing at B=10T (~15.5nm), therefore any inter-vortex coupling is expected to be weak here (see Part-III of SM for more discussions, which includes Ref. [26]). Below we will focus our study on free vortex cores.

As the double-gapped superconducting spectrum was commonly observed [12, 18], we shall examine its origin. For single-layer FeSe/$SrTiO_3$, there are two electron pockets on each M point in the folded Brillouin Zone (BZ) (Fig. 1(e)), and there could be finite band hybridization between them (Fig. 1(f), on which the nodeless $d$-wave pairing relies [21]). So far various ARPES studies have discovered significant gap anisotropy on *single* electron pocket, however hybridizations have not been observed within the experimental resolution [17,19]. Here we find that the ARPES measured anisotropic gap function in Ref. 19:

$$\Delta_k = \Delta_0 - A\cos(2\theta_k) + B\cos(4\theta_k) \qquad \text{------------------ [1]}$$

could account for the double-gapped spectrum. As illustrated in Fig. 1(g), such a gap function produces two local gap maxima of $\Delta_2$ (=$\Delta_0$+$A$+$B$) at $\theta_k$ = π/2, and 3π/2, and $\Delta_1$ (=$\Delta_0$−$A$+$B$) at $\theta_k$ =0, and π, which generate two pairs of coherence peaks in dI/dV; while $\Delta_0$ is the mean gap over the Fermi surface. The corresponding fit in Fig. 1(c) yields $\Delta_0$ = 10.58meV, $A$ = 3.25meV and $B$ = 2.87meV. Details of the fitting procedure are described in Part-IV of SM.

Fig. 2 presents the tunneling spectra of four different free vortices, obtained at $T$=0.4K ($T_{elec}$ =1.18K). Fig. 2(a) shows the dI/dV line cut taken across Vortex 1, with clear multiple discrete peaks near the core center. These peaks locate symmetrically with respect to $E_F$, but *no* ZBCP is observed. Fig. 2(e) shows the spatial evolution of the spectra in a color plot. Discrete states can be seen within a ±2nm range around the center and vanish outside; meanwhile, a pair of broader peaks show up at higher energies (shaded regions in Fig. 2(a)). Those broader peaks keep moving to high energy and eventually merge into the coherence peaks. Similar behaviors were observed on other free vortices (see Fig. 2(b) for Vortex 2 and Fig. S6 for Vortices 3,4).

To better resolve the low-energy core states, Fig. 2(c) and 2(d) show the dI/dV focusing on small energy scale (±6 meV), taken at the cores of Vortices 3 and 4. Clearly, there are up to six well separated peaks symmetrically distributed with respect to $E_F$. Those peaks are equally spaced and their positions almost keep the same within ±1.2 nm around the core. A color plot of Fig. 2(c) is shown in Fig. 2(f).

We applied multiple-Gaussians fitting to the summed spectra near the center of Vortices 1-4, as shown in Figs. 3(a-d). The fitted peak energies are labeled, and detailed fitting parameters including the peak width and fitting errors are presented in Part-VI of SM. Fig. 3(e) shows the normalized peak energies of vortices 1-4 by dividing with their averaged spacing ($\delta E$). The results sit well on the lines of *half* integer value (±1/2, ±3/2 or ±5/2), which is expected for the CdGM states of an *s*-wave superconductor. Moreover, the high-energy shifting peaks are expected to be closely-packed states, as the spacing of CdGM states decreases at high energy; and their maximal intensity locations move away from the core center [4]. Therefore, we resolved both discrete low-energy states and quasi-continuous high-energy states in single-layer FeSe, for its large gap, small $E_F$, and high resolution here.

Nonetheless, the energy spacing $\delta E$ varies from 1.1 meV to 1.9 meV for different vortices, which is likely due to superconducting gap variations. To have a quantitative analysis, we extracted the mean size ($\Delta_0$) of the local gap where vortices 1~4 emerge (Fig. S2), by fitting them with the same way as discussed in Part-IV of SM. We found that the $\delta E$ of different vortices can be reasonably accounted by $(\Delta_0)^2/E_F$ (taking a constant $E_F$ = 60 meV from our previous QPI study [18]). A linear fit of $(a\Delta_0)^2/E_F$ to $\delta E$ yields $a$ = 0.95(±0.14) (Fig. 3(f)). Therefore, a single anisotropy gap can account for both superconducting gap spectrum and the CdGM states, band hybridization is not necessarily involved here.

The behaviors of the CdGM states of single-layer FeSe/$SrTiO_3$ match those of an *s*-wave superconductor. The absence of ZBCP here excludes topological superconductivity as in Fe(Se,Te) [6,7] and (Li,Fe)OHFeSe [8,9], or chiral *p*-wave pairing. Nonetheless, a theoretical study suggests that a nodeless *d*-wave pairing is possible if the two electron pockets of FeSe/$SrTiO_3$ are hybridized by SOC [21], as sketched in Fig. 1(f). Since finite SOC has been widely observed in iron-based superconductors [27], and in particular ARPES study [19] sets the upper limit of the SOC of FeSe/$SrTiO_3$ to be 5 meV (limited by its resolution), we shall examine whether this scenario could explain our observation. We simulated the vortex states of 1ML FeSe/$SrTiO_3$ based on a two-band $\boldsymbol{k}\cdot\boldsymbol{p}$ model that includes SOC and hybridization [20], under both *s*-wave and nodeless *d*-wave pairing. Details of the model are described in Part-VII of SM, and in which the parameters are chosen to mimic experimentally measured Fermi surface and superconducting gap.

Fig. 4(a) presents the simulation for plain *s*-wave pairing at zero SOC strength ($\lambda=0$). It displays typical CdGM states that symmetrically distributed around $E_F$ with equal spacing. As approaching the core center, the intensities of low level states increase and show certain particle-hole asymmetry [4]. These overall behaviors qualitatively agree with our data in Fig. 2 and Fig. S6. After applying SOC up to $\lambda=0.03t$ (≈ 4meV, $t$ = 135meV is a model parameter), there is no obvious change in the simulated CdGM state, as demonstrated in Fig. 4(c) and Fig. S9. This is because for *s*-wave pairing, SOC simply shifts the chemical potential of the two hybridized bands (to opposite direction), the resulting modification on CdGM state ($E = \mu\Delta^2/(E_F \pm \lambda)$) is negligible when $\lambda << E_F$.

For nodeless *d*-wave pairing, finite SOC must be present to avoid band crossing (Fig. 1(f)). Fig. 4(b) shows the corresponding simulation with $\lambda$=0.02t (~2.7meV). Noticeably, it displays *two* sets of CdGM states (from the two hybridized bands), with energies shifted to opposite directions. We further found such energy shift is of the similar amount of $\lambda$, as shown in Fig. 4(d). A simple understanding is that for nodeless *d*-wave which relies on hybridization, the SOC acts as a shift of "chemical potential" for Bogoliubov-de Gennes (BdG) quasiparticles [21], which will directly shift vortex states as $E = (\mu\Delta^2/E_F) \pm \lambda$ (see Part-VII of SM for more discussion). However, this contradicts our experimental observations of symmetrically distributed and equally spaced CdGM states within an uncertainty ≤ ±0.02 meV (see Tab. S2), despite the sizable variations of the gap and $\delta E$. In fact, any asymmetric offset or splitting larger than a fraction of our resolution (0.36 meV) would have been easily resolved by STS. As SOC is a necessary condition (and in Ref. 21 sizable SOC is needed to explain the ARPES observed gap anisotropy), our observation thus strongly disfavors nodeless *d*-wave pairing.

We note that despite the energy of CdGM states well agree with *s*-wave pairing, their intensities show some irregular distribution around the core center, as seen in Figs. 2(e,f). Further simulation (Part-VIII of SM) show that this could be due to randomly distributed interfacial disorders in FeSe/$SrTiO_3$, which can locally affect the intensity of CdGM state. We also note the Zeeman effect at B =10T is unlikely to affect the superconducting pairing here, since the Zeeman energy (~1.1 meV) is still significantly smaller the gap of FeSe/$SrTiO_3$ (~10meV).

To gain more information on the pairing, we examined the states at the boundaries of FeSe/$SrTiO_3$ film. It was suggested that for a $d_{x2-y2}$-wave superconductor, zero-energy ABS would exist at the {[110]} oriented boundaries due to the $\pi$-phase shift in the reflection of quasi-particles [10]; while topologically non-trivial superconductors possess dispersive ABS at all surface/edges, which give finite DOS within the superconducting gap [11]. However, usually ABS is not expected at the boundary of conventional s-wave superconductors.

In the FeSe film presented in Fig. 1(a), there are boundaries between 2ML FeSe and 1 ML FeSe regions with *continues* surface lattice. The non-superconducting 2ML FeSe [12] made such boundaries well-defined 1D S/N interfaces. Fig. 5(a) shows such a boundary along $\boldsymbol{a_0}$, i.e. the [110] direction of the Fe lattice, where zero-energy ABS is expected for *d*-wave pairing with sign-change between adjacent M points (Fig. 1(f)). Fig. 5(c) shows a dI/dV line cut taken across such a boundary, with a spatial interval of 0.6 nm (marked in Fig. 5(a)). The gap of 1ML FeSe, with flat and zero-DOS bottom, keeps untouched until very close to interface (≤1 nm);

then the gap quickly disappeared at the interface and shows a metallic DOS on the 2ML FeSe side. There is no ZBCP formed at both sides of the interface.

Fig. 5(b) presents another type of boundary: the [110] oriented step edge of 1ML FeSe. The dI/dV spectra taken along the edge, within a distance ≤ 1nm, are shown in Fig. 5(d). The majority of the spectra show a full superconducting gap with flat bottom; while some spectra occasionally show irregular in-gap states (e.g. spectra 1, 2, 5, 6). Since a discontinues step edge is more easily to have local disorders and defects, the fully gapped spectrum is likely the intrinsic feature on the step edge while the in-gap states are generated by local disorders. Overall, there is no zero-energy ABS or intrinsic in-gap states on both types of [110] oriented boundaries, which disfavors *d*-wave or topologically non-trivial superconductors (e.g., chiral *p*-wave), but is consistent with the plain *s*-wave pairing.

Our studies of CdGM states and boundary states provide independent evidences on the *s*-wave pairing in single-layer FeSe/$SrTiO_3$. This helps to clarify the current controversy on its pairing symmetry, and is consistent with recently proposed cooperative pairing enhancement scenario [16, 28]. Moreover, our results provide insight for the fast-developing field of topological superconductivity in iron-based superconductors. Interlayer coupling has been shown to create band inversion and topological surface states in Fe(Se,Te) [23, 24] and (Li,Fe)OHFeSe [8], which eventually leads to possible Majorana zero modes. For FeSe/$SrTiO_3$, a 2D system, the absence of ZBCP in vortex core and the absence of in-gap features at 1D boundaries in our data show that it is likely topologically trivial, although recent studies show that there is a SOC-induced gap below $E_F$ at M points [22, 29]. Therefore, our results further suggest that interlayer coupling is a prerequisite for topological superconductivity of iron-based superconductors.

**Acknowledgements:**

We thank Professors Jiangping Hu and Yihua Wang for helpful discussions. This work is supported by the National Natural Science Foundation of China (Grant Nos.: 11888101, 11790312, 11421404, 11574134), National Key R&D Program of the MOST of China (Grant Nos.: 2016YFA0300200, 2016YFA0300401, 2017YFA0303004), Science Challenge Project (grant No. TZ2016004).

**References:**

[1]. C. Caroli, P. G. de Gennes, and J. Matricon, Bound Fermion states on a vortex line in a type II superconductor. J. Phys. Lett. **9**, 307 (1964).

[2]. G. E. Volovik, Fermions on quantized vortices in superfluids and superconductors. Turk. J. Phys. **20**, 693 (1996).

[3]. H. F. Hess, et al. Scanning-tunneling-microscope observation of the Abrikosov flux lattice and the density of states near and inside a fluxoid. Phys. Rev. Lett. **62**, 214 (1989).
[4]. N. Hayashi, T. Isoshima, M. Ichioka, and K. Machida, Low-lying quasiparticle excitations around a vortex core in quantum limit. Phys. Rev. Lett. **80**, 2921 (1998).
[5]. M. Y. Chen, et al. Discrete energy levels of Caroli-de Gennes-Matricon states in quantum limit in $FeTe_{0.55}Se_{0.45}$. Nat. Commun. **9**, 970 (2018).
[6]. D. F. Wang, et al. Evidence for Majorana bound state in an iron-based superconductor. Science **362**, 333-335 (2018).
[7]. T. Machida, et al. Zero-energy vortex bound state in the superconducting topological surface state of Fe(Se,Te) Nature Materials **18**, 811 (2019)
[8]. Q. Liu, et al. Robust and clean Majorana zero mode in the vortex core of high-temperature superconductor $(Li_{0.84}Fe_{0.16})OHFeSe$, Phys. Rev. X **8**, 041056 (2018)
[9]. C. Chen, et al., Quantized conductance of Majorana zero mode in the vortex of the topological superconductor (Li0.84Fe0.16)OHFeSe. Chin. Phys. Lett. **36**, 057403 (2019)
[10]. Satoshi, K. & Yukio, T. Tunnelling effects on surface bound states in unconventional superconductors. Rep. Prog. Phys. **63** 1641–1724 (2000)
[11]. X. L. Qi and S. C. Zhang, Topological Insulators and Superconductors. Rev. Mod. Phys. **83**, 1057 (2011).
[12]. Wang, Q. Y. et al. Interface induced high temperature superconductivity in single unit-cell FeSe films on $SrTiO_3$. Chin. Phys. Lett. **29**, 037402 (2012).
[13]. Liu, D. F. et al. Electronic origin of high-temperature superconductivity in single-layer FeSe superconductor. Nature Commun. **3**, 931 (2012).
[14]. He, S. L. et al. Phase diagram and electronic indication of high-temperature superconductivity at 65K in single-layer FeSe films. Nature Mater. **12**, 605–610 (2013).
[15]. Tan, S. Y. et al. Interface-induced superconductivity and strain-dependent spin density waves in $FeSe/SrTiO_3$ thin films. Nature Mater. **12**, 634–640 (2013).
[16]. Lee, J. J. et al. Interfacial mode coupling as the origin of the enhancement of $T_c$ in FeSe films on $SrTiO_3$. Nature **515**, 245-248 (2014).
[17]. Peng, R. et al. Measurement of an enhanced superconducting phase and a pronounced anisotropy of the energy gap of a strained FeSe single layer in $FeSe/Nb{:}SrTiO_3/KTaO_3$ heterostructures using photoemission spectroscopy. Phys. Rev. Lett. 112, 107001 (2014).
[18]. Q. Fan, et al. Plain s-wave superconductivity in single-layer FeSe on $SrTiO_3$ probed by scanning tunneling microscopy, Nature Physics, **11**, 946 (2015)
[19]. Zhang, Y., et al. Superconducting Gap Anisotropy in Monolayer FeSe Thin Film. Phys. Rev. Lett. **117**, 117001 (2016).
[20]. Hirschfeld, P. J. et al. Gap symmetry and structure of Fe-based superconductors. Rep. Prog. Phys. **74** 124508 (2011).
[21]. Agterberg, D. F. et al. Resilient Nodeless d-Wave Superconductivity in Monolayer FeSe. Phys. Rev. Lett. **119**, 267001 (2017).
[22]. N. Hao and J. Hu, Topological phase in the single-layer FeSe, Phys. Rev. X, **4**, 031053 (2014).
[23]. Z. Wang, et al. Topological nature of the $FeSe_{0.5}Te_{0.5}$ superconductor, Phys. Rev. B. **92**, 115119 (2015).
[24]. G. Xu, et al. Topological superconductivity on the surface of Fe-based superconductors. Phys. Rev. Lett. **117**, 047001 (2016).

[25]. See Supplemental Material for additional data, method and discussions.
[26]. M. Cheng, R. M. Lutchyn, V. Galitski, S. Das Sarma, Splitting of Majorana-Fermion Modes due to Intervortex Tunneling in a px + ipy Superconductor. Phys. Rev. Lett. **103**, 107001 (2009).
[27]. S.V. Borisenko, et al., Direct observation of spin–orbit coupling in iron-based superconductors, Nature Physics, **12**, 311 (2016)
[28]. Q. Song et al. Evidence of cooperative effect on the enhanced superconducting transition temperature at the $FeSe/SrTiO_3$ interface, Nature Communications, **10**, 758 (2019)
[29]. Z. F. Wang, et al. Topological edge states in a high-temperature superconductor $FeSe/SrTiO_3(001)$ film, Nature Materials, **15**, 968–973 (2016)

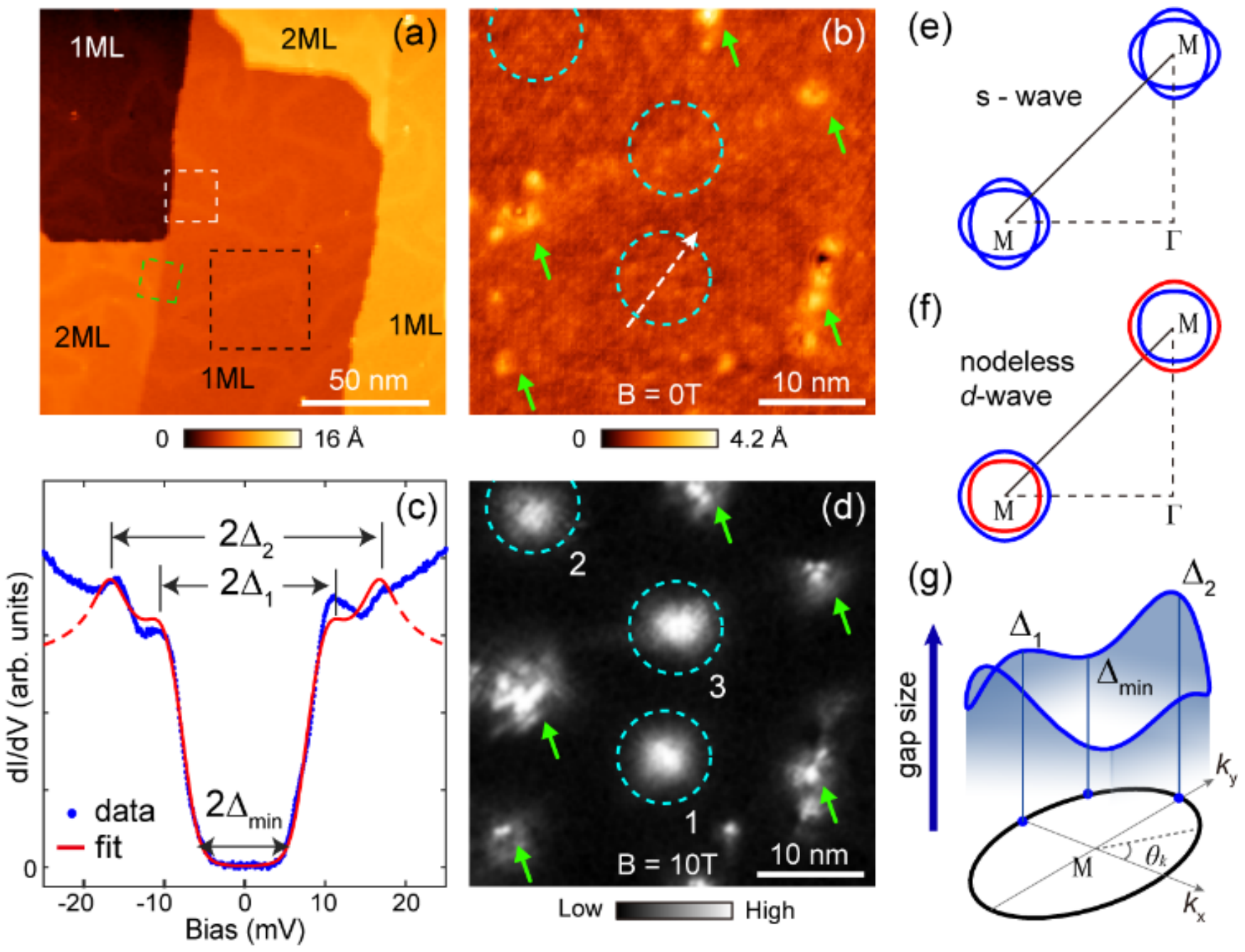

**Fig. 1.** (**a**) STM image of $FeSe/SrTiO_3$ film with a thickness of ~1.3 ML (160×160 nm$^2$, $V_b$=3V, $I$=5pA). (**b**) Image of a 1ML region (40×40 nm$^2$) (**c**) Averaged gap spectrum of 1ML FeSe ($V_b$=30mV, I=60pA, T=4.2K), taken along the white arrow in (b) at B = 0T, and the gap fitting using Eq. 1. (**d**) Zero-bias dI/dV mapping taken on the same area of (b), at B = 10T. Green arrows in (b) and (d) indicate surface defects and the pinned vortices, dashed circles indicate free vortices. (**e**) and (**f**): Sketch of Fermi surface of 1ML FeSe in the folded BZ with $s_{(++)}$ wave and nodeless $d$-wave pairing, respectively. (**g**) Sketch of the gap distribution on the electron pocket at M (determined by Eq.1). $\Delta_2$, $\Delta_1$ and $\Delta_{min}$ correspond to the two local gap maxima and the gap minima, respectively, and $tg(\theta_k) = k_y/k_x$.

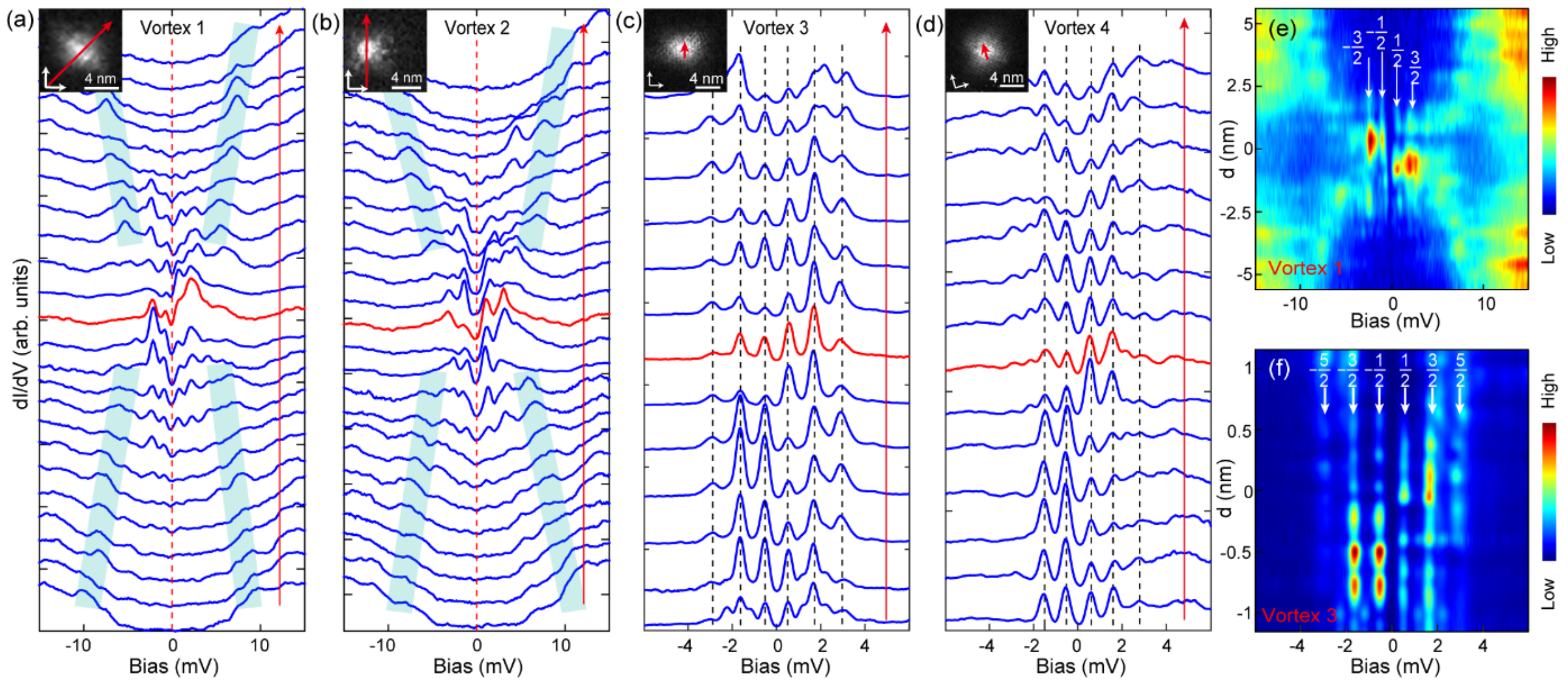

**Fig. 2.** Discrete bound states in free vortices. (**a**)-(**b**) dI/dV spectra taken across Vortex 1 and 2 ($V_b$ = 20mV, $I$ = 60pA, $\Delta V$ = 0.1mV). (**c**)-(**d**) dI/dV spectra taken in small energy range across the center of vortices 3 and 4 ($V_b$ = 6 mV, $I$ = 60 pA, $\Delta V$ = 0.1mV). Red curves in (a~d) are collected at the vortex center (estimated), and insets are vortex maps. (**e**)-(**f**) Color plot of the spectra in (a) and (c), respectively. Arrows indicate the individual CdGM states. All the spectra are taken at T = 0.4K ($T_{elec}$ = 1.18K)

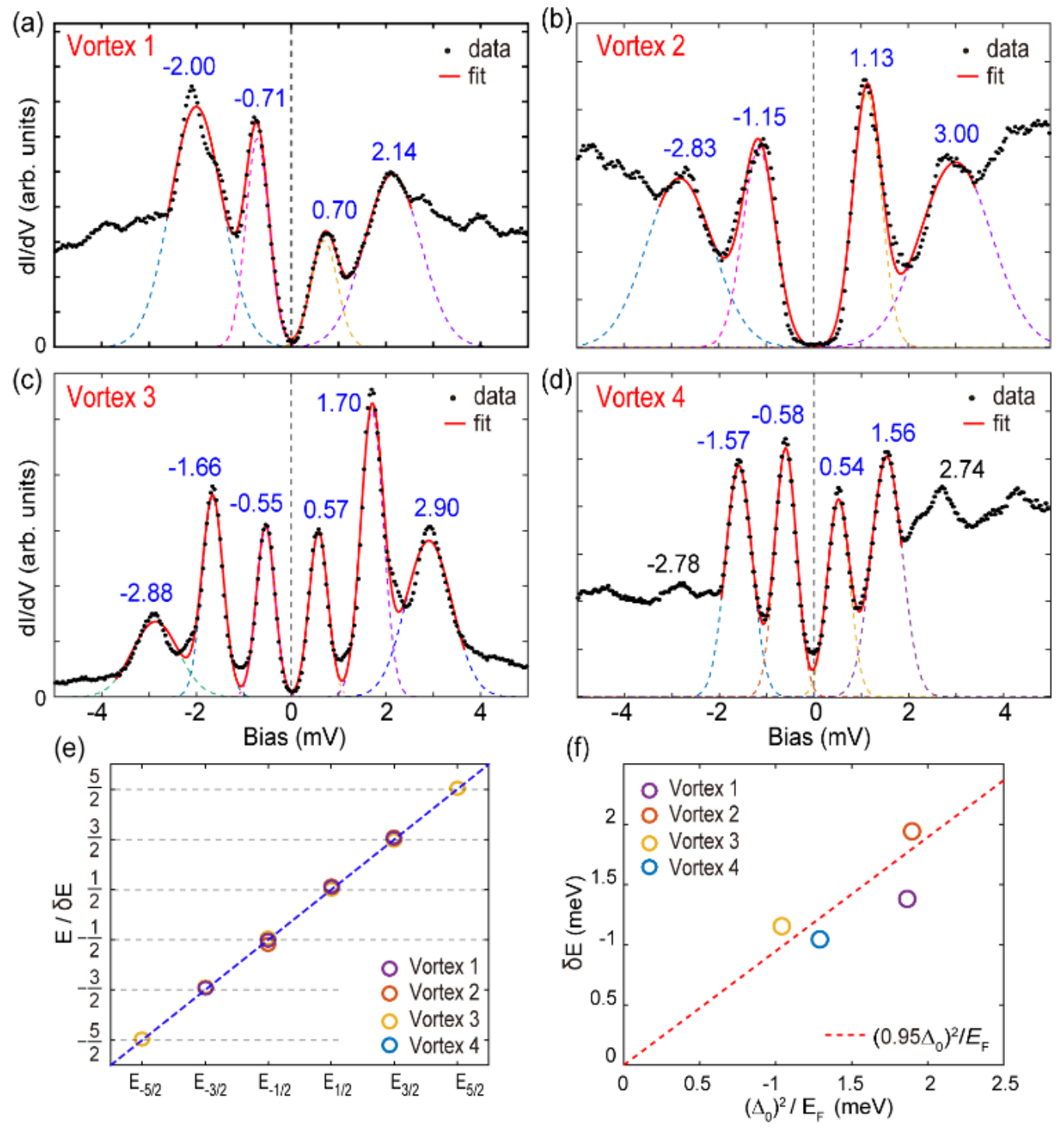

**Fig. 3.** Quantitative fitting of vortex bound states. (**a-d**) Low energy spectra of free Vortices 1-4. Red curves are multiple Gaussian-peak fits (Dashed curves are individual peaks). (**e**) Normalized energy of the CdGM state of Vortices 1-4, via dividing the averaged δE of each vortex. (**f**) The relation of $(\Delta_0)^2/E_F$ and $\delta E$ for Vortices 1-4, dashed line is the linear fitting.

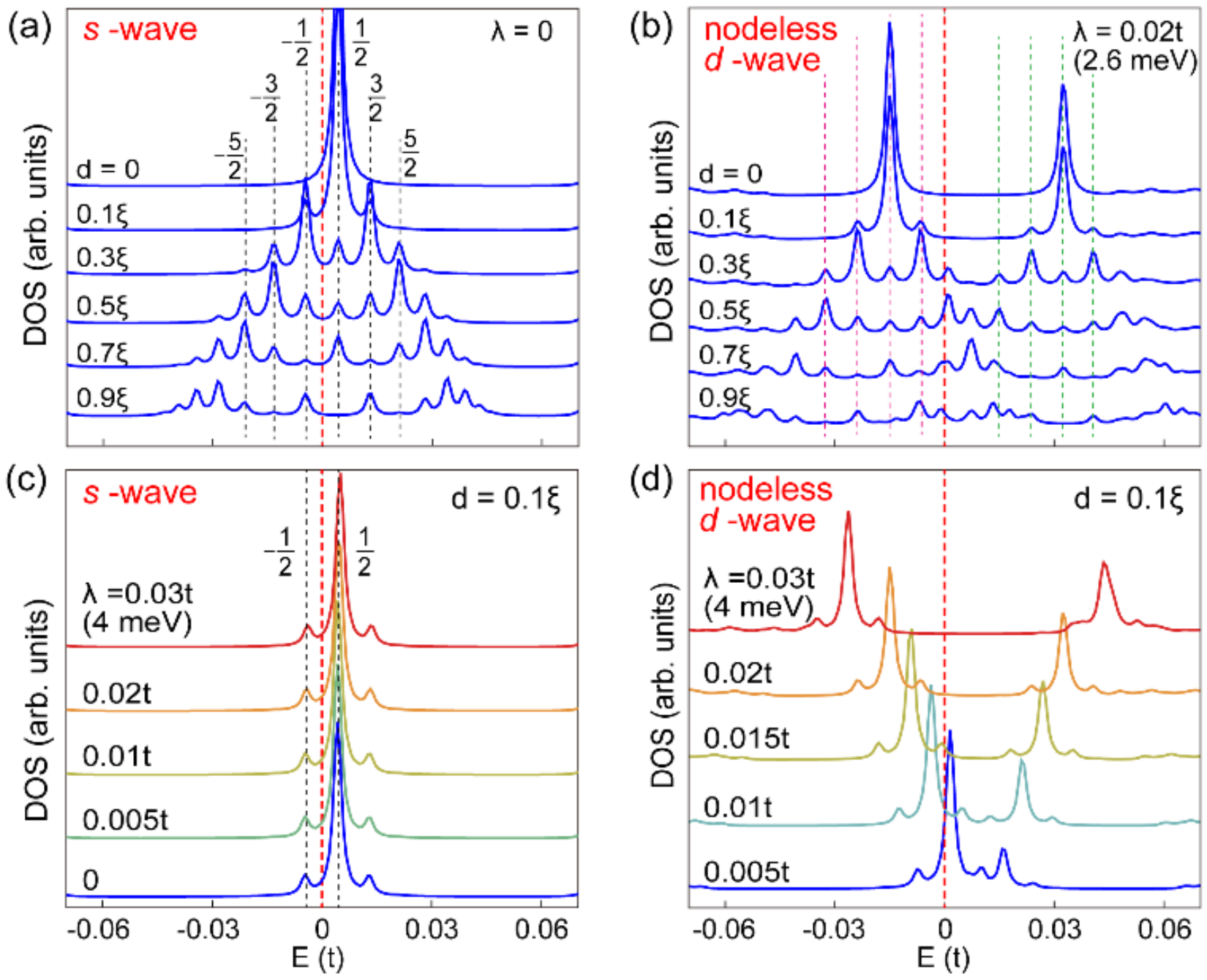

**Fig. 4.** Calculated vortex states under *s*-wave pairing (**a**) without SOC, and nodeless *d*-wave pairing (**b**) with $\lambda = 0.02$t, at different distance to the core center ($\xi$ is the coherence length). Calculated vortex states under *s*-wave (**c**) and nodeless *d*-wave (**d**) pairing at d=0.1$\xi$, with various SOC strength.

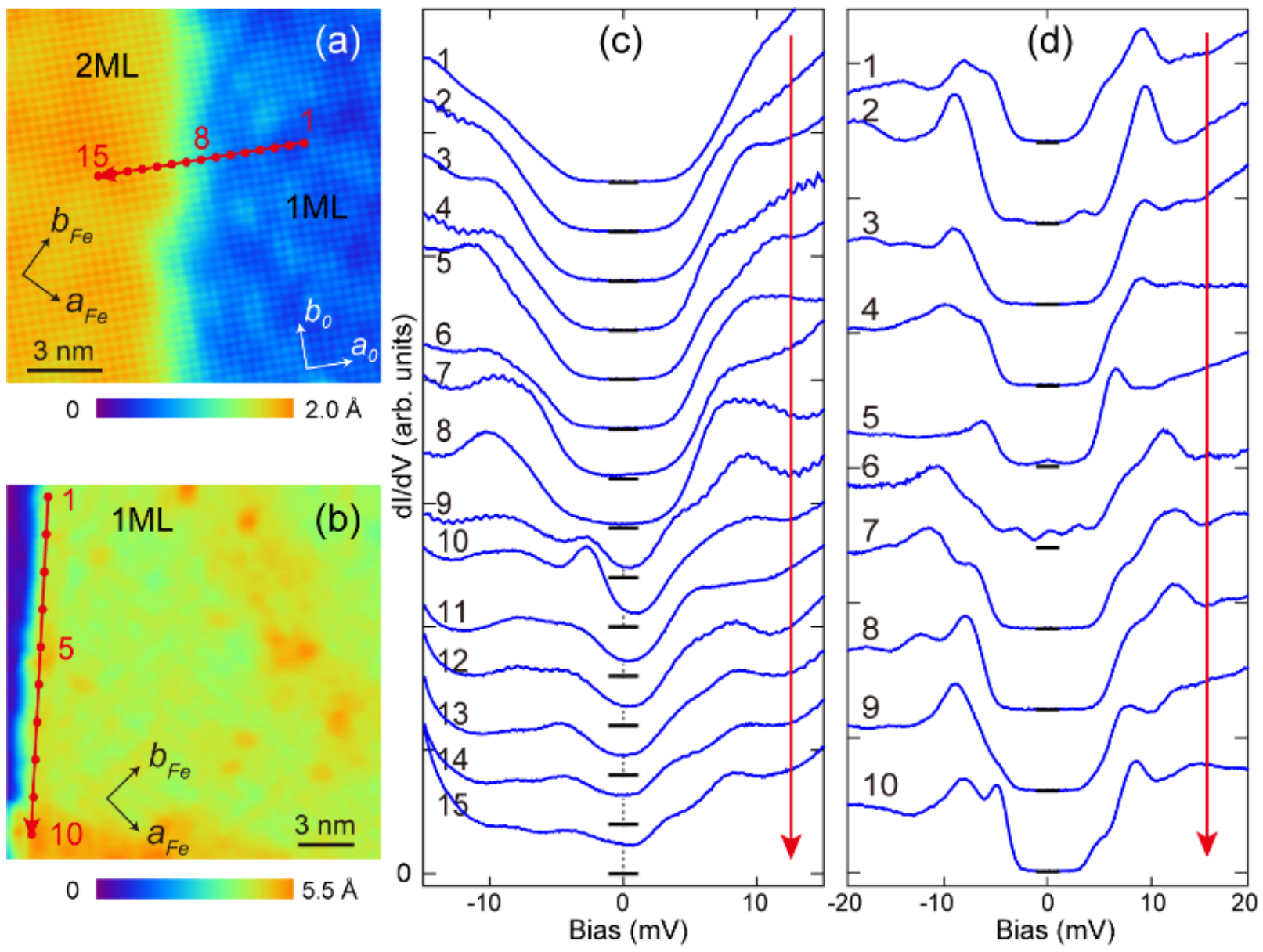

**Fig. 5.** STM images of a boundary between 2ML/1ML FeSe (**a**) and atomic step edge of 1ML FeSe (**b**), taken within the green and white dashed square in Fig. 1(a), respectively. **(c-d)**. dI/dV spectra taken along the arrow in panel (a) and (b), respectively (T = 4.2K). The short black bars indicate the zero-conductance position of each curve.

# Supplementary Materials of "Observation of discrete conventional Caroli-de Gennes-Matricon states in the vortex core of single-layer $FeSe/SrTiO_3$"

Chen Chen[1,4†], Qin Liu[1,2,4†], Wei-Cheng Bao[3,6†], Yajun Yan[1,4], Qiang-Hua Wang[3,4*], Tong Zhang[1,4*], Donglai Feng[1,4,5*]

[1] State Key Laboratory of Surface Physics, Department of Physics, and Advanced Materials Laboratory, Fudan University, Shanghai 200438, China

[2] Science and Technology on Surface Physics and Chemistry Laboratory, Mianyang, Sichuan 621908, China

[3] National Laboratory of Solid State Microstructures & School of Physics, Nanjing University, Nanjing, 210093, China

[4] Collaborative Innovation Center of Advanced Microstructures, Nanjing 210093, China

[5] Hefei National Laboratory for Physical Science at Microscale and Department of Physics, University of Science and Technology of China, Hefei, Anhui 230026, China

[6] Zhejiang University of Water Resources and Electric Power, Hangzhou 310018, China

[†]These authors contributed equally.

## I. Calibration of STM energy resolution at T = 0.4 K and bias voltage offset.

The energy resolution of a low-T STM is limited by thermal and electrical noise broadening. It can be estimated by $3.5k_BT_{elec}$ where $T_{elec}$ is the effective electron temperature. To calibrate $T_{elec}$, we measured the superconducting gap of a Pb/Si(111) film at $T$ = 0.4 K, as shown in Fig. S1(a). A standard BCS fit gives $\Delta$ = 1.39 meV, $T_{elec}$ = 1.18 K and a small Dynes term $\Gamma$ = 0.005 meV that accounts for finite quasi-particle lifetime. The energy resolution of the STM is then given by $3.5k_BT_{elec}$ = 0.36 meV.

The STM bias voltage ($V_b$) applied to the sample usually has a small offset. This offset will affect the determination of the energy position of the CdGM states, and thus should be carefully corrected. The actual zero-point of $V_b$ can be calibrated by measuring $I$-$V$ curves at different setpoint current, because they intersect at a single point where $V$ = 0 and $I$ = 0. Fig. S1(b) shows such $I$-$V$ measurement at a free vortex core and the zero-point of $V_b$ can be determined with a precision of ± 0.01 meV. Fig. S1(c) show the dI/dV measured at the same position. Clearly, the CdGM states are distributed symmetrically with respect to $E_F$ (with a precision of ± 0.02 meV), and there is no ZBCP. All the dI/dV curves throughout the paper are calibrated by the same way.

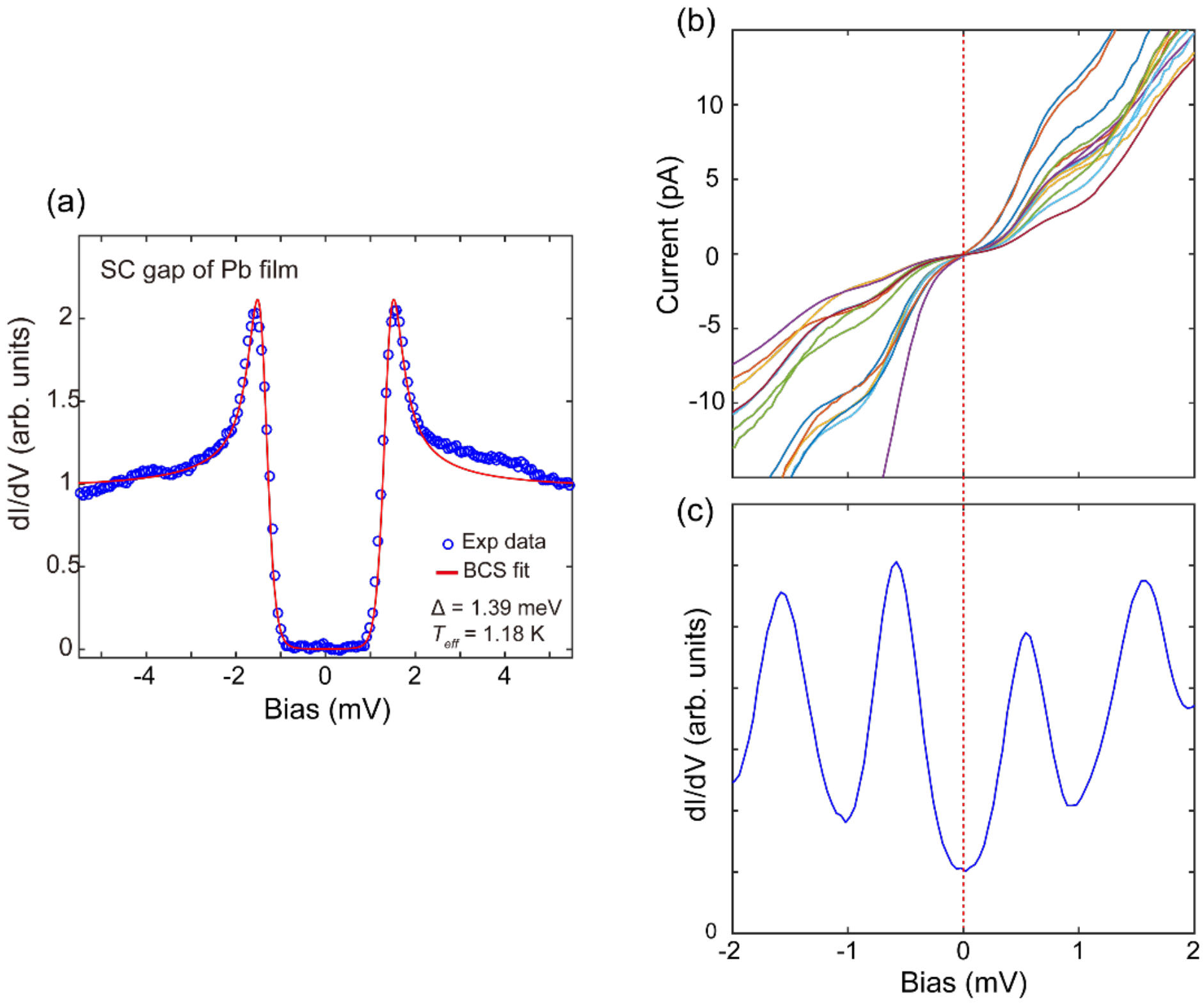

**Fig. S1.** (**a**) The superconducting gap of the Pb/Si(111) film (blue dashed circles) measured at $T$ = 0.4 K ($T_{elec}$=1.18K). The red curve is the BCS fit with $\Delta$ = 1.39 meV, $T_{elec}$ = 1.18 K and $\Gamma$ = 0.005 meV. (**b**) A set of *I-V* spectra taken at different setpoints on a free vortex core of 1ML FeSe/STO. The zero point of $V_b$ is determined by their crossing point. (**c**) dI/dV spectrum taken at the same tip position as (b), which shows symmetrically distributed CdGM states around $E_F$, without ZBCP.

## II. Additional data of local superconducting gap measured at free vortex regions and near the surface defects.

The superconducting gap spectra (dI/dV line cut) taken at zero field, across the area where Vortex 1~4 appear when magnetic field applied, are shown in Fig. S2(a~d). The spatially averaged gap spectra are shown in Fig. S2(e).

The STM image and dI/dV spectra taken around typical surface defects in 1ML FeSe/$SrTiO_3$ are shown in Fig. S3. The defects appear as irregular bumps in the image and in most cases there are in-gap states observed near the defect site. The nature of these surface defects is unclear for now (they could be impurities from the MBE source or from the substrate). Upon applying magnetic field, these defects can easily pin vortex cores, in which the CdGM states is strongly affected by impurity state, as reflected by the rather irregular shaped vortex core marked by arrows in Fig. 1(d). To avoid the influence of the defects state, we only study the vortices which display clean local superconducting gap at zero field.

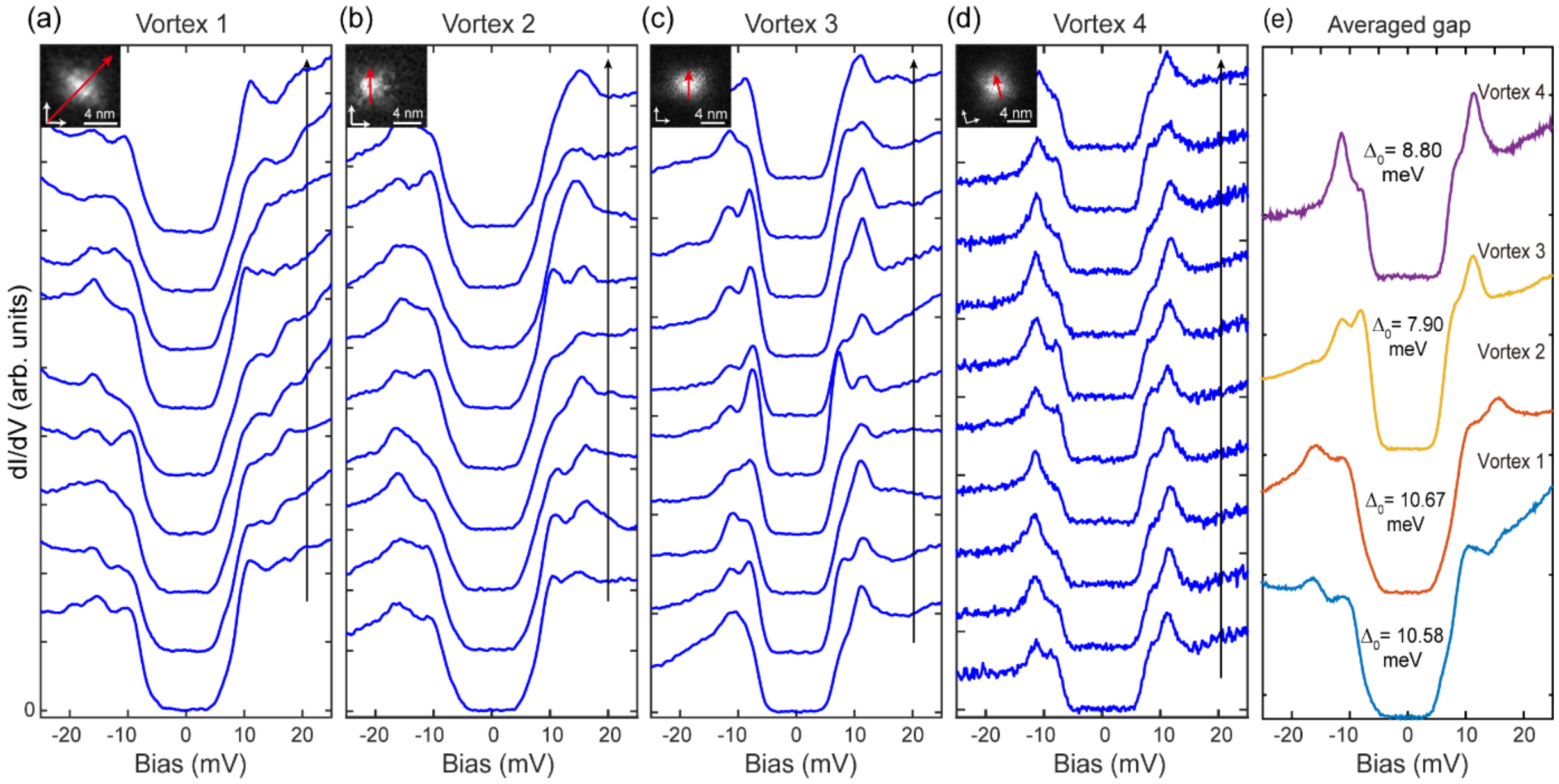

**Fig. S2** (**a~d**) Superconducting gap spectra taken across the area where Vortex 1~4 appear (along the arrows in the inset images), respectively. Panel (a~c) are measured at T=4.2K while panel (d) is measured at T=0.4K ($T_{elec}$=1.18K). (**e**) Averaged spectra from the dI/dV line cut shown in (a~d), for Vortex 1~4. The mean gap sizes ($\Delta_0$) are obtained from the gap fitting shown below.

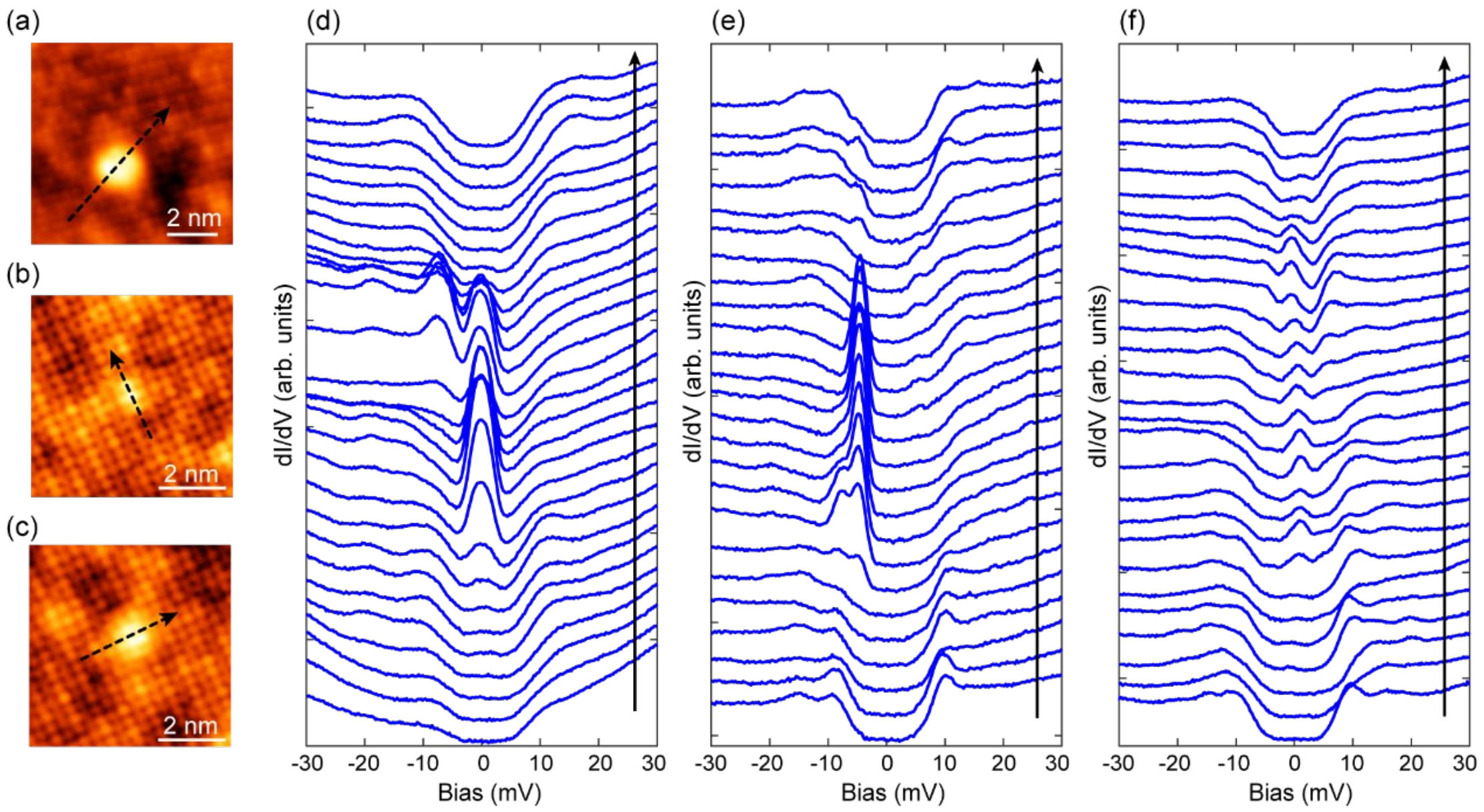

**Fig. S3.** Impurity state induced by native defects in 1ML FeSe/$SrTiO_3$. (**a, b, c**) Topographic images of typical surface defects in 1ML FeSe/$SrTiO_3$ film. (**d, e, f**) dI/dV spectra taken along the arrows in panel (a), (b) and (c), respectively (measured at $T$ = 4.2K). Pronounced in-gap states can be observed near the defect site.

## III. Measurement of the coherence length of 1 ML $FeSe/SrTiO_3$

The coherence length of $FeSe/SrTiO_3$ can be estimated by the spatial decay of low-energy vortex state near $E_F$. Fig. S4(b) shows an averaged zero-bias map of a single vortex (measured at 4.2K). Exponential fits to its line profile yields $\xi$ of 2.03 nm and 2.45 nm along the [110] and [100] directions, respectively. The slight anisotropy of $\xi$ is likely due the anisotropy of the superconducting gap and/or Fermi velocity. (Note that at T = 4.2K, discrete CdGM state cannot be resolved and they merge into a broad zero-bias peak at core center (ref. 18). Thus the decay of zero-bias conductance can be used to estimate coherence length).

It is known that if the inter-vortex spacing is close to $\xi$, the inter-vortex coupling will make the vortex state or Majorana zero mode hybridize and split (ref. S1). At B = 10T, assuming a triangular vortex lattice, the averaged vortex spacing (d) is expected to be ~15.5nm. This is consistent with our measurement as shown Fig. S4(a), in which the distance between vortex 1 and 2, vortex 2 and 3 is 13nm and 17.4nm, respectively. Thus the inter-vortex spacing is still much larger than $\xi$ at B=10T ($d/\xi > 6$ in average), so the core-core interaction should be sufficiently weak (according to ref. [26], the vortex interaction also decays as $\exp(-d/\xi)$).

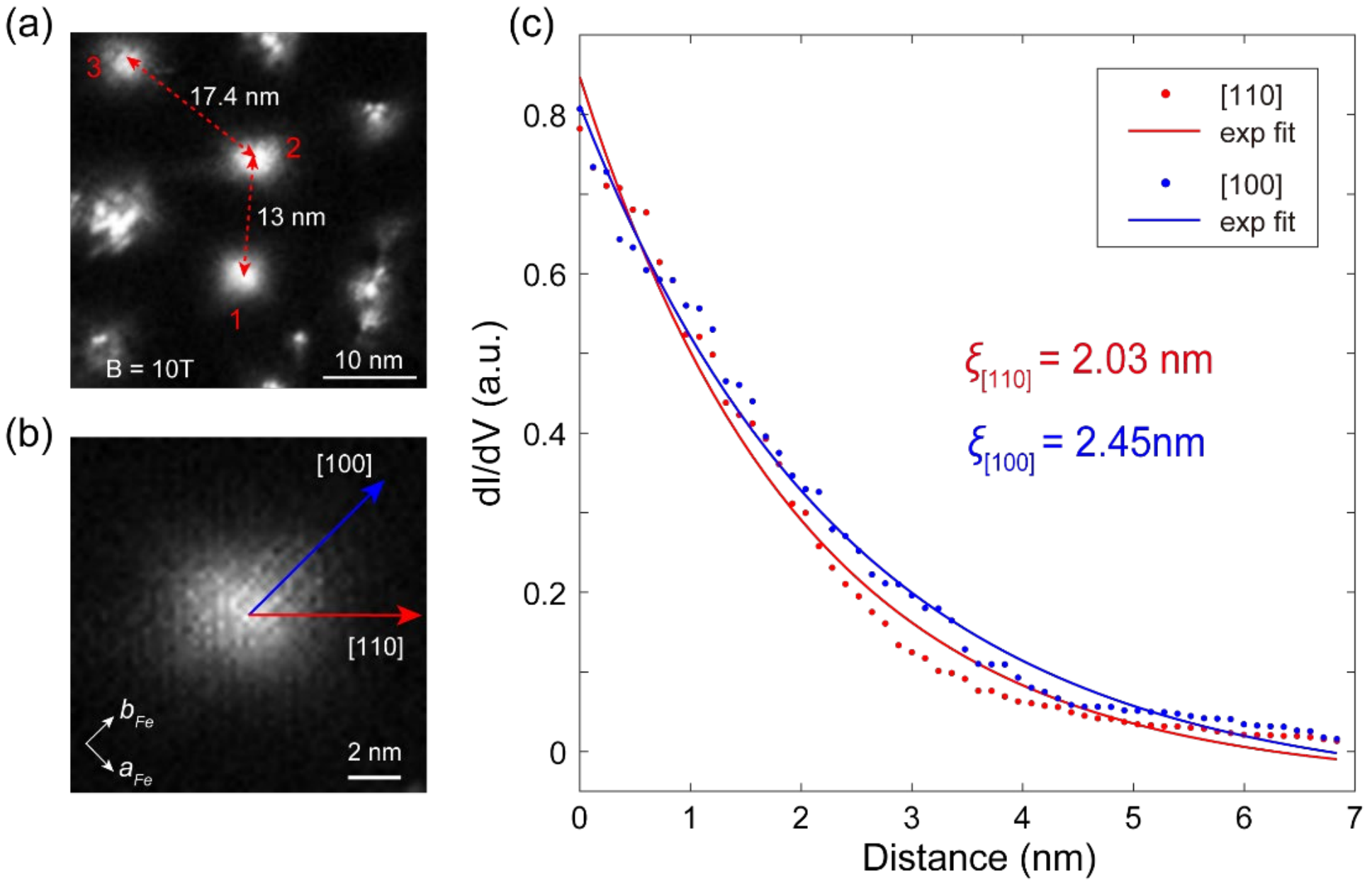

**Fig. S4**. (a) The same vortex mapping as that shown in Fig. 1(d) (T = 4.2K, ΔV = 1 mV). The distance between vortex 1 and 2, vortex 2 and 3 is marked. (b) Zero-bias dI/dV mapping of a single vortex core (averaged from vortex 1~2 in (a)), the orientation of the Fe lattice is marked. (c) Dots: vortex line profile taken along the [110] (red) and [100] (blue) directions shown (b); Solid lines: exponential fits to the line profile ($y \propto \exp(-d/\xi)$), which yield the $\xi$ of 2.03 nm and 2.45 nm along [110] and [100] directions, respectively.

## IV. Fitting the superconducting gap of 1 ML FeSe/SrTiO3

According to the ARPES measurement in Ref. 18, the superconducting gap distribution on a single electron pocket of 1 ML FeSe/SrTiO3 can be described by an anisotropic gap function:

$$\Delta_k = \Delta_0 - A\cos(2\theta_k) + B\cos(4\theta_k)$$

where $\Delta_0$, $A$ and $B$ are positive parameters, and $\mathrm{tg}(\theta_k) = k_y/k_x$. Such gap function will produce two local gap maxima of $\Delta_2$ ($=\Delta_0+A+B$) and $\Delta_1$ ($=\Delta_0-A+B$) at $\theta_k = \pi/2$ and $\theta_k = 0$ (respectively), and a gap minimum of $\Delta_{\min}$ ($=\Delta_0-B$) at $\theta_k = \pi/4$, as sketched in Fig. 1(g). The mean gap size over the Fermi surface: $\overline{\Delta} = \frac{1}{2\pi}\int_0^{2\pi} \Delta_k \, d\theta_k$, is equal to $\Delta_0$. We use this gap function to fit the measured tunneling spectrum. The superconducting DOS is given by the Dynes formula:

$$N(E)_k = \left|\mathrm{Re}\left(\frac{E - i\Gamma}{\sqrt{(E - i\Gamma)^2 - \Delta_k^2}}\right)\right|$$

where $\Gamma$ is a broadening factor due to finite quasi-particle lifetime. The total tunneling conductance is then given by:

$$\frac{dI}{dV} \propto \int N(E)_k f'(E+eV)\, dk\, dE$$

where $f'(E)$ is the derivative of Fermi-Dirac function at an effective temperature ($T'$) which also accounts for the broadening of the spectrum.

The measured gap spectrum usually has a sloping background, as shown in Fig. 1(c) and Fig. S2. The DOS at positive energy is always higher. Such particle-hole asymmetry is possible due to the relatively shallow band (or small $E_F$) of 1 ML FeSe/SrTiO3. Therefore before fitting, we divided the original $dI/dV$ by a linear background to reduce the line-shape asymmetry, e.g., as that shown in Figs. S5(a~b) for the gap of vortex 1. Then gap fitting is applied by using above formula (red curve in Fig. S5(b)). Since the simulated gap is symmetric with respect to $E_F$, it better matches the symmetrized $dI/dV$ spectrum (obtained by averaging the dI/dV values at +/- bias), as shown in Fig. S5(c). In Figs. S5(d~f), we show the slope corrected and symmetrized gap spectrum of Vortices 2~4 and the corresponding gap fittings. The detailed fitting parameters are listed in Tab. S1.

It is seen that the fitted values of gap size and broadening factors are different for different vortices. This could be due to the variation of local doping and unevenness of the interface in FeSe/SrTiO3. As the $\Gamma$ values are fairly small, the gap broadening is actually controlled by $T'$ (given by $3.5k_BT'$). In the last column of Tab. S1 we list the estimated gap broadening (To compare with the broadening of the CdGM states measured at 0.4K, a 3.8K thermo-broadening is deducted from the gap broadening of Vortices 1~3, as them are measure at 4.2K. The gap of Vortex 4 is measured at 0.4K).

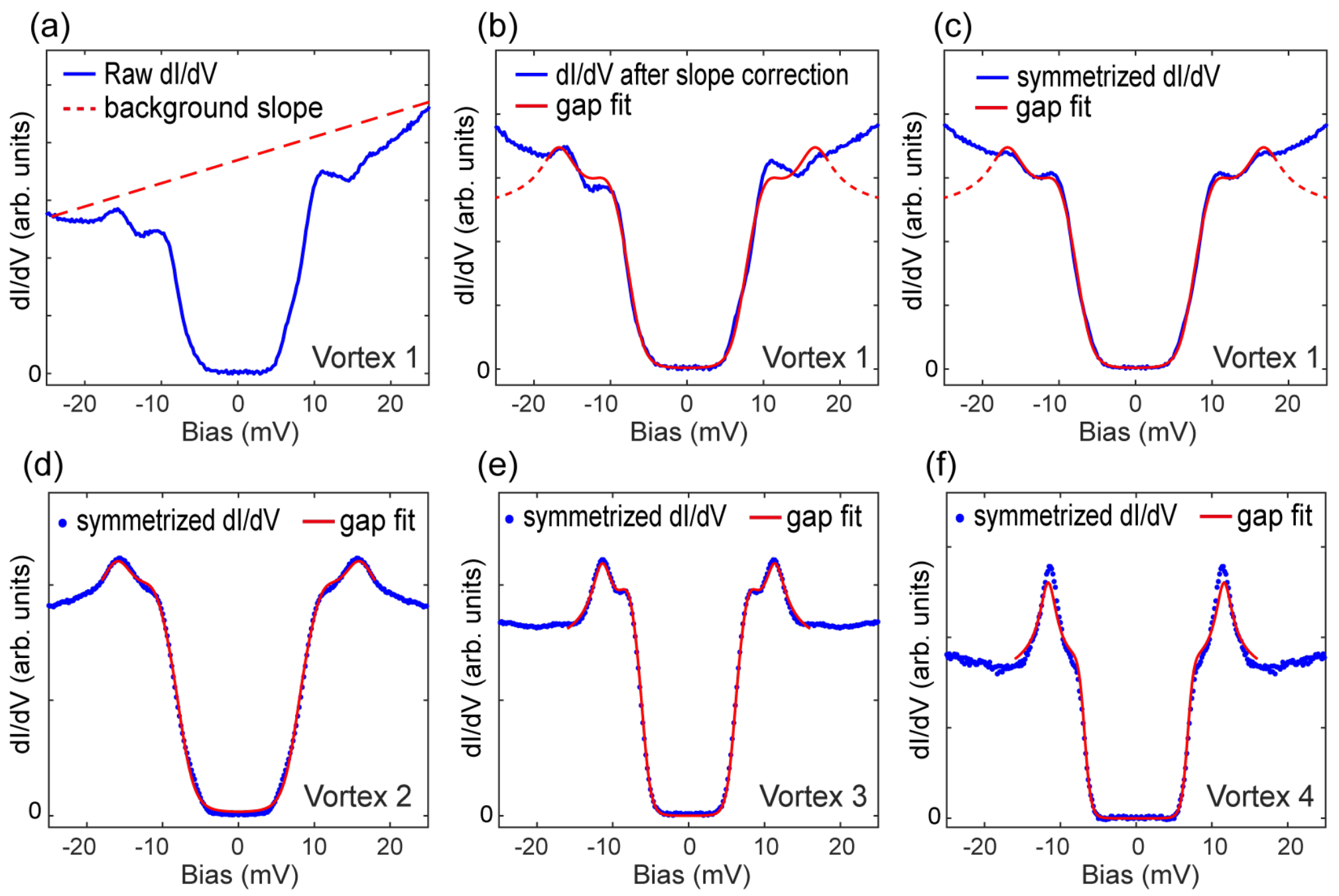

**Fig. S5 | Fittings of the local superconducting gap measured on Vortices 1~4.** (**a**) The (averaged) dI/dV spectrum taken near the location of Vortex 1 (at B=0T). Red dashed line is the background slope defined by a linear fit to gap shoulders. (**b**) The gap spectrum of Vortex 1 after slope correction. Red curve is the gap fitting. (**c**) Symmetrized gap spectrum of Vortex 1 with respect to $E_F$ (after slope correction). The fitting is the same as that in panel (b). (**d~f**) Symmetrized gap spectra of vortices 2~4 after slope correction (raw spectra are shown Fig. S2(e)), respectively, and the corresponding gap fittings.

| **Vortex** | $\Delta_0$ | $A$ | $B$ | $\Delta_1$ | $\Delta_2$ | $T'$ (K) | $\Gamma$ | Broadening of SC gap* |
|---|---|---|---|---|---|---|---|---|
| **1** | 10.58 ± 0.11 | 3.25 | 2.87 | 10.2 | 16.7 | 10.42 | 0.026 | 1.99 |
| **2** | 10.67 ± 0.05 | 2.53 | 2.86 | 11.0 | 16.06 | 14.2 | 0.011 | 3.14 |
| **3** | 7.90 ± 0.05 | 1.67 | 1.71 | 7.94 | 11.28 | 5.66 | 0.077 | 0.56 |
| **4** | 8.80 ± 0.22 | 2.37 | 0.50 | 6.93 | 11.67 | 4.95 | 0.015 | 1.49 |

**Table. S1** (Unit: meV, except for $T'$): Fitting parameters of the local superconducting gap of Vortex 1~4.
* For Vortices 1~3, the SC gap broadening is calculated by $3.5k_B(T'-3.8)$, while for Vortex 4 it is calculated by $3.5k_BT'$.

## V. Additional data of the CdGM state in free vortex cores.

In Fig. S6 we show the topographic image and dI/dV mapping of Vortex 4, and additional dI/dV line cut data taken across Vortex 3 and 4. In Fig. S7 we show the dI/dV line cut taken along different orientation across Vortex 1.

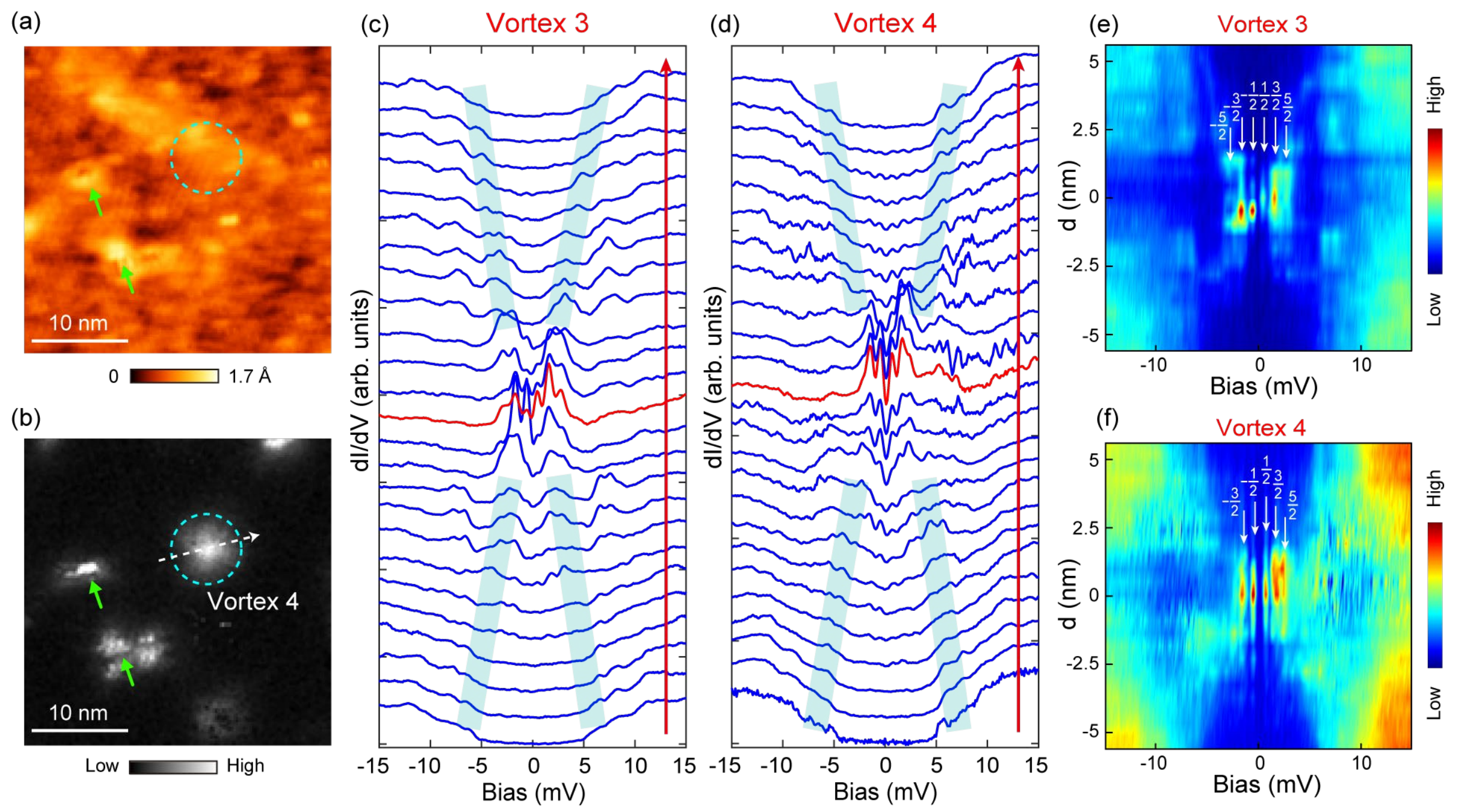

**Fig. S6.** Additional data of vortex state measurement. (**a, b**) Topographic image and zero-bias dI/dV mapping of the region where Vortex 4 located. The circle indicates Vortex 4 and the arrows indicate surface defects. (**c, d**) dI/dV line cut (large energy range) taken across Vortices 3 and 4, respectively. (**e, f**) Color plots of (c) and (d), respectively.

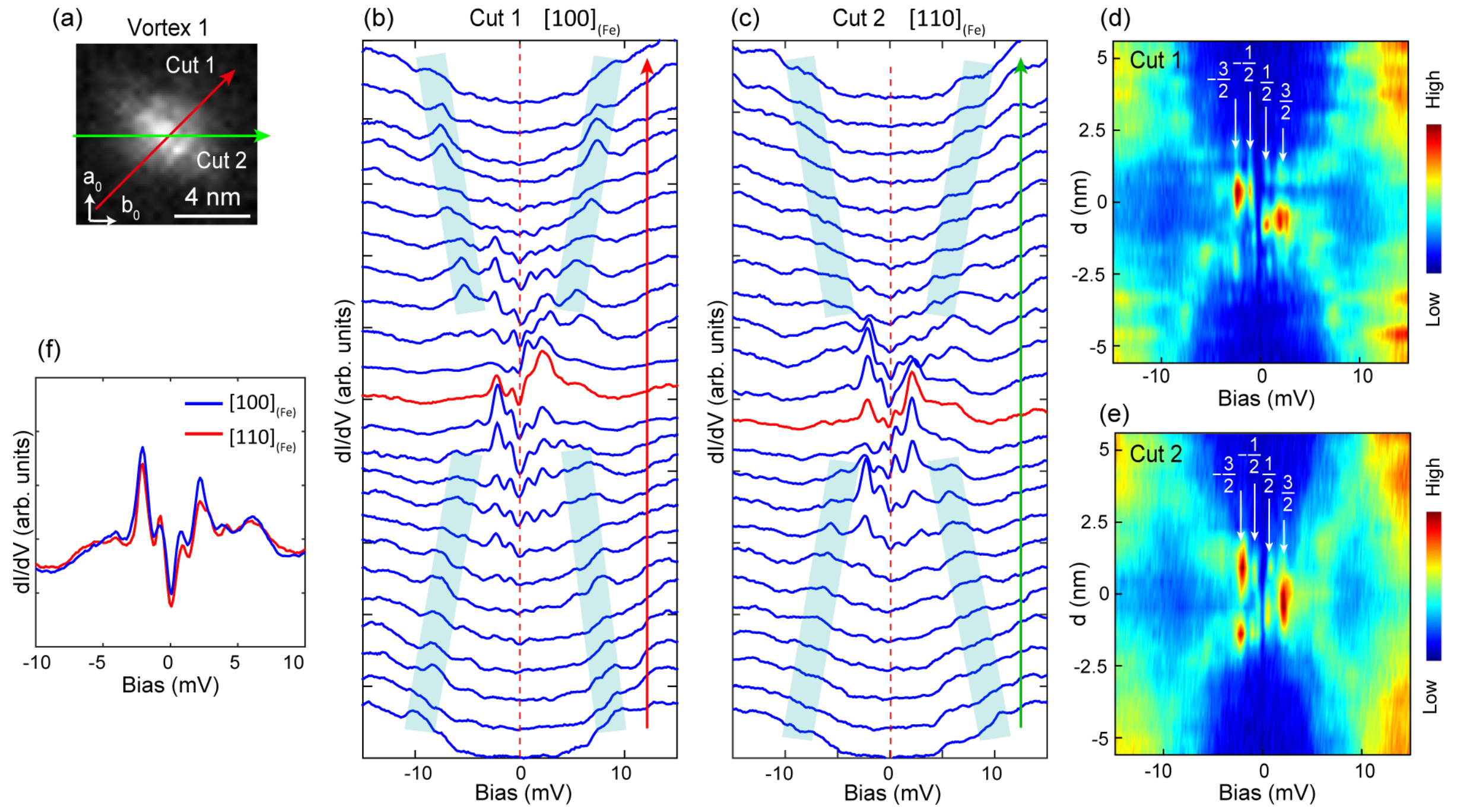

**Fig. S7.** Additional line cut data of Vortex 1. (**a**) dI/dV mapping of Vortex 1. (**b, c**) dI/dV line cuts taken along $[100]_{Fe}$ direction (red arrow) and $[110]_{Fe}$ direction (green arrow) in (a). (**d, e**) Color plot of (b) and (c), respectively. (**f**) Averaged STS over a ±2 nm range of the line cut in (b) and (c).

## VI. Multiple Gaussian function fitting of CdGM bound states

To give quantitative analysis on the CdGM states, we fit the summed spectra near the centers of Vortex 1~4 with multiple Gaussian peaks. The fitting curves are shown in Figs. 3(a~d) and the fit parameters are summarized Table S2, including the peak energy and peak width (FWHM). The fitting error (95% confidence bound) are listed following the main value. Within our resolution, the fitted peak energies locate symmetrically with respect to $E_F$. In the last column of Table S2, we show the degree of "asymmetry" of the $E_{\pm 1/2}$ peaks which have the smallest fitting error. The asymmetry is defined by $(E_{1/2}+E_{-1/2})/2$ and is <=0.02 meV for all the vortices. In Table S3 we listed the measured energy spacing (δE) of CdGM states and the estimated values by their local SC gap ($\Delta_0$).

We note the fitted peak width (full-width at half-maximum or FWHM) is in the range of 0.46 - 0.81 meV for $E_{\pm 1/2}$ states and 0.52 - 1.8 meV for $E_{\pm 3/2}$ states (see Tab. S2), which are still larger than the energy resolution here (0.36 meV). It could be partially due to the gap anisotropy as discussed above, and partially due to the quasi-particle scattering effects, as the low-energy states are always sharper. In Table S3 we also listed the gap size, gap broadening and the FWHM of CdGM states for all the vortices. It is seen that there is a positive correlation between the broadening of SC gap and the broadening of CdGM peaks, which suggests they could be affect by the same factor, such as the disorder scattering from the interface.

| Vortex | | $E_{-5/2}$ | $E_{-3/2}$ | $E_{-1/2}$ | $E_{1/2}$ | $E_{3/2}$ | $E_{5/2}$ | $(E_{-1/2}+E_{1/2})/2$ |
|---|---|---|---|---|---|---|---|---|
| 1 | **Energy** | | -2.00 ± 0.10 | -0.71 ± 0.01 | 0.70 ± 0.02 | 2.14 ± 0.04 | | -0.005 ± 0.01 |
| | **FWHM** | | 1.30 ± 0.04 | 0.58 ± 0.01 | 0.65 ± 0.03 | 1.40 ± 0.08 | | |
| 2 | **Energy** | | -2.83 ± 0.08 | -1.15 ± 0.02 | 1.13 ± 0.01 | 3.00 ± 0.10 | | -0.01 ± 0.015 |
| | **FWHM** | | 1.65 ± 0.16 | 0.81 ± 0.03 | 0.70 ± 0.02 | 1.80 ± 0.14 | | |
| 3 | **Energy** | -2.88 ± 0.05 | -1.66 ± 0.01 | -0.55 ± 0.01 | 0.57 ± 0.01 | 1.70 ± 0.01 | 2.90 ± 0.02 | 0.01 ± 0.01 |
| | **FWHM** | 1.15 ± 0.09 | 0.52 ±0.02 | 0.46 ± 0.02 | 0.47 ± 0.02 | 0.56 ± 0.01 | 1.14 ± 0.04 | |
| 4 | **Energy** | -2.78* | -1.57 ± 0.01 | -0.58 ± 0.01 | 0.54 ± 0.01 | 1.56 ± 0.01 | 2.74* | -0.02 ± 0.01 |
| | **FWHM** | | 0.66 ± 0.02 | 0.54 ± 0.01 | 0.56 ± 0.01 | 0.77 ± 0.02 | | |

**Table. S2** (Unit: meV): Fitting parameters of the core state peaks of Vortex 1~4.

* Peak energies measured directly from the dI/dV curve, where fitting is not applicable.

| Vortex | $\overline{\delta E}$ | $\Delta_0$ | $\frac{(0.95\Delta_0)^2}{E_F}$ | Broadening of SC gap | FWHM of $E_{\pm 1/2}$ * | FWHM of $E_{\pm 3/2}$ ** |
|---|---|---|---|---|---|---|
| 1 | 1.38 | 10.58 | 1.68 | 1.99 | 0.61 | 1.35 |
| 2 | 1.94 | 10.67 | 1.71 | 3.14 | 0.75 | 1.72 |
| 3 | 1.15 | 7.90 | 0.94 | 0.56 | 0.46 | 0.54 |
| 4 | 1.05 | 8.80 | 1.16 | 1.49 | 0.55 | 0.72 |

**Table. S3** (Unit: meV): Averaged spacing of CdGM states and related SC gap parameters for Vortex 1~4.
* mean value of the FWHM of $E_{-1/2}$ and $E_{1/2}$ states.
** mean value of the FWHM of $E_{-3/2}$ and $E_{3/2}$ states.

## VII. Vortex states simulated by a two-band model of 1 ML FeSe/$SrTiO_3$

In this section we simulate the vortex core states of 1ML FeSe/$SrTiO_3$ under both nodeless *d*-wave pairing and plain s-wave pairing, using a two-band model. The electronic structure of 1ML FeSe/$SrTiO_3$ has two electron pockets at *M*. Before they are folded into the reduced 2Fe/cell Brillouin zone, they can be viewed as a pocket around X and another around Y in the 1Fe/cell Brillouin zone. Upon folding, the two pockets intersect, as shown in Fig. S8(a). Following Ref. 21 we adopt a ***k·p*** model and compactify it on a lattice. This is sufficient to describe the low energy quasi-particles. The normal-state single-particle Hamiltonian is, in momentum space,

$$h_{\boldsymbol{k}}=\epsilon_{\boldsymbol{k}}+d_{\boldsymbol{k}}\sigma_3+\boldsymbol{g}_{\boldsymbol{k}}\cdot\boldsymbol{s}\,\sigma_1$$

Henceforth, $\sigma_{1,2,3}$ are Pauli matrices acting on the two effective orbitals, and $s_{1,2,3}$ are Pauli matrices acting on spins. Agterburg *et al.* proposed that in the continuum limit [21]:

$$\epsilon_{\boldsymbol{k}}=\frac{k^2}{2m}-E_f,\quad d_{\boldsymbol{k}}=\alpha k_x k_y,\quad \boldsymbol{g}_{\boldsymbol{k}}=\beta(k_y,\,k_x)$$

where *m* is the effective mass, and $\alpha$ and $\beta$ are coefficients. The above form of $h_{\boldsymbol{k}}$ takes proper account of the symmetry of the effective orbitals at the *M* point near the Fermi level. We rotate the coordinate frame (by 45° about *z*) and spin axis independently, so that

$$\epsilon_{\boldsymbol{k}}\rightarrow\frac{k^2}{2m}-E_f,\quad d_k\rightarrow\alpha'(k_x^2-k_y^2),\quad \boldsymbol{g}_{\boldsymbol{k}}\rightarrow\beta'(k_y,\,k_x)$$

with modified coefficients. Notice that rotation of the spin axis by constant Euler angles does not alter the singlet pairing. The advantage of the rotated $h_{\boldsymbol{k}}$ is an emergent superficial $\boldsymbol{C_{4v}}$ symmetry, with the two orbitals behaving effectively as *xz* and *yz*. (Notice that the resulting Fermi pockets are elongated along *x* and *y* in the new frame.) We then compactify the model on a lattice, with

$$\epsilon_{\boldsymbol{k}}\rightarrow-2t(\cos k_x+\cos k_y)-\mu,\quad d_{\boldsymbol{k}}\rightarrow2t'(\cos k_x-\cos k_y),\quad \boldsymbol{g}_{\boldsymbol{k}}\rightarrow2\lambda(\sin k_y,\sin k_x)$$

where *t'* accounts for the Fermi pocket anisotropy, and $\lambda$ measures the strength of the spin-orbital coupling (SOC). We set $t=1$, $\mu=-3.63t$ (or $E_F=0.37t$) and $t'=0.125t$ to have shallow electron pockets mimicking the experimental situation, as shown in Fig. S8(c~d). By comparing the Fermi energy of model and that of experiment, we can think $t = 1$ roughly corresponds to 135meV. Ref. 19 sets the upper limit of the SOC of 1 ML $FeSe/SrTiO_3$ to be 5 meV (limited by its resolution). Therefore in the calculation we set $\lambda \leq 0.03t$ (~4 meV). A further advantage of the rotated $h_{\boldsymbol{k}}$ arises after the compactification: the symmetric hopping integrals, the *d*-wave like anisotropy and the SOC can all be defined on nearest-neighbor bonds.

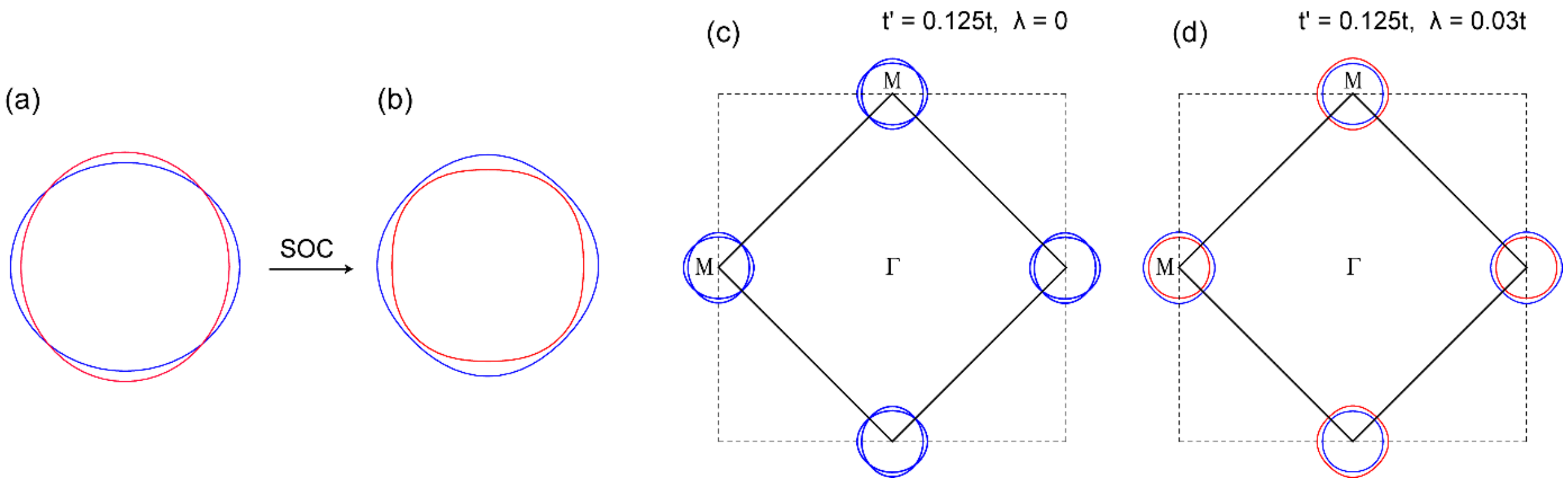

**FIG. S8.** (**a**, **b**) Schematic of electron-like Fermi pockets around the *M* point in the Brillouin zone, in the absence (a) and presence (b) of SOC, respectively. The color in (a) shows the nodeless *d*-wave gap function which is positive (red) on one elliptic pocket and negative (blue) on the other. After including SOC, these pockets are reconstructed into the inner (red) and outer (blue) pockets in (b). (**c, d**) Sketch of the Fermi surface of $\boldsymbol{k}\cdot\boldsymbol{p}$ model obtained from the normal state Hamiltonian with assigned parameters $t$, $t'$ and $\mu$ (see the text), and $\lambda = 0$ for (c) and $\lambda = 0.03t$ for (d). SOC induced hybridization can be clearly seen.

We first consider s-wave pairing case. In the superconducting state with *s*-wave pairing, the vortex bound states would be the usual Caroli-de Gennes-Matricon states. We write the pairing part of the Hamiltonian in momentum space as

$$\Delta_{\boldsymbol{k}} = \Delta_0 \sigma_0 i s_2$$

where $\Delta_0$ is the onsite part same for the two bands before reconstruction, $is_2$ is the spin antisymmetric tensor accounting for singlet pairing, and we set $\Delta_0 = 0.07t$ (~10 meV) for the simulation. The simulated LDOS at different distance to the vortex center, and with different SOC strength ($\lambda$), are shown in Figs. S9(a~f). One can see all the panels displays symmetrically distributed CdGM states around $E_F$, which do not change notably with SOC. A qualitative understanding is that under s-wave pairing, although SOC induces band hybridization, it only slightly shifts the chemical potential of the two hybridized bands to opposite direction ($E_F' = E_F \pm \lambda$), while the superconducting gap size does not change. The resulting change in the CdGM state energy $E = \mu\Delta^2/(E_F \pm \lambda)$ is too small to be detected when $\lambda << E_F$.

Then we consider the nodeless *d*-wave pairing proposed in Ref. 21. In such a case, there is a full gap on both pockets, but with a sign change. One may worry that upon hybridization, the energy gap on the reconstructed bands, illustrated in Fig. S8(b), would have to be nodal.

However, this does not have to be so if the hybridization is from SOC, as shown in Ref. 21. We now ask how the nodeless $d$-wave would impact on the vortex bound states. We write the pairing part of the Hamiltonian as, in momentum space (for the uniform case),

$$\Delta_{\boldsymbol{k}}=[\Delta_1\sigma_3+\Delta_2(\cos k_x-\cos k_y)]\ is_2$$

where $\Delta_1$ is the onsite part, with opposite phase on the two bands before reconstruction, see Fig. S8(a) for illustration, $\Delta_2$ is the amplitude of $d$-wave pairing on nearest bonds, and $is_2$ is the spin antisymmetric tensor accounting for singlet pairing (note the definitions of $\Delta_1$ and $\Delta_2$ here are different from those in the main text). Note that $\sigma_3$ transforms as $d$-wave under rotation, hence both components in the gap function behave as $d$-wave. The resulting pairing gap is nodeless if $|\Delta_1|>4|\Delta_2|$. Since $\Delta_2$ leads to gap variation on the Fermi surface, which is weak in experiments (about 20% from ARPES), we ignore $\Delta_2$ for the moment and set $\Delta_1= 0.07t$ (we verified that including this part does not alter the results qualitatively). By writing $\Delta_{\boldsymbol{k}}$ in real space with the non-uniform pairing in a vortex state, we can calculate the local density of states (LDOS) along a line cut approaching the vortex core.

In Fig. S10(a~f) we show the simulation for nodeless $d$-wave with $\lambda$ varies from $0.005t$ to $0.03t$. Remarkably, they all display *two* sets of CdGM state with energies shifted away from $E_F$ (towards opposite direction), and the energy shift is of similar amount of SOC. Phenomenologically, this can be understood as that for nodeless $d$-wave pairing, SOC directly enter the Bogoliubov-deGennes(BdG) equation which determines the quasiparticle dispersion [21]. It acts as a shift of "chemical potential" for BdG quasiparticles and then directly shifts the peak position of the vortex states, as $E = (\mu\Delta^2/E_F) \pm \lambda$ (where "+" and "-" sign apply to two different bands, respectively). Compare to the case of s-wave pairing, such energy shift or splitting is much more significant. However, in STM study we always observe symmetrically distributed and equally spaced CdGM states, despite the sizable variations of the local SC gap and energy spacing ($\delta E$). Such two sets of shifted CdGM states as that shown in Fig. S10 has never been observed. In fact, any offset or splitting of the CdGM peak larger than a fraction of our resolution (0.36 meV) would have been easily resolved by STS. Since SOC is necessary for nodeless $d$-wave pairing, and according to Ref. 21 sizable SOC is needed to explain the ARPES observed gap anisotropy, our observation thus disfavor this possibility.

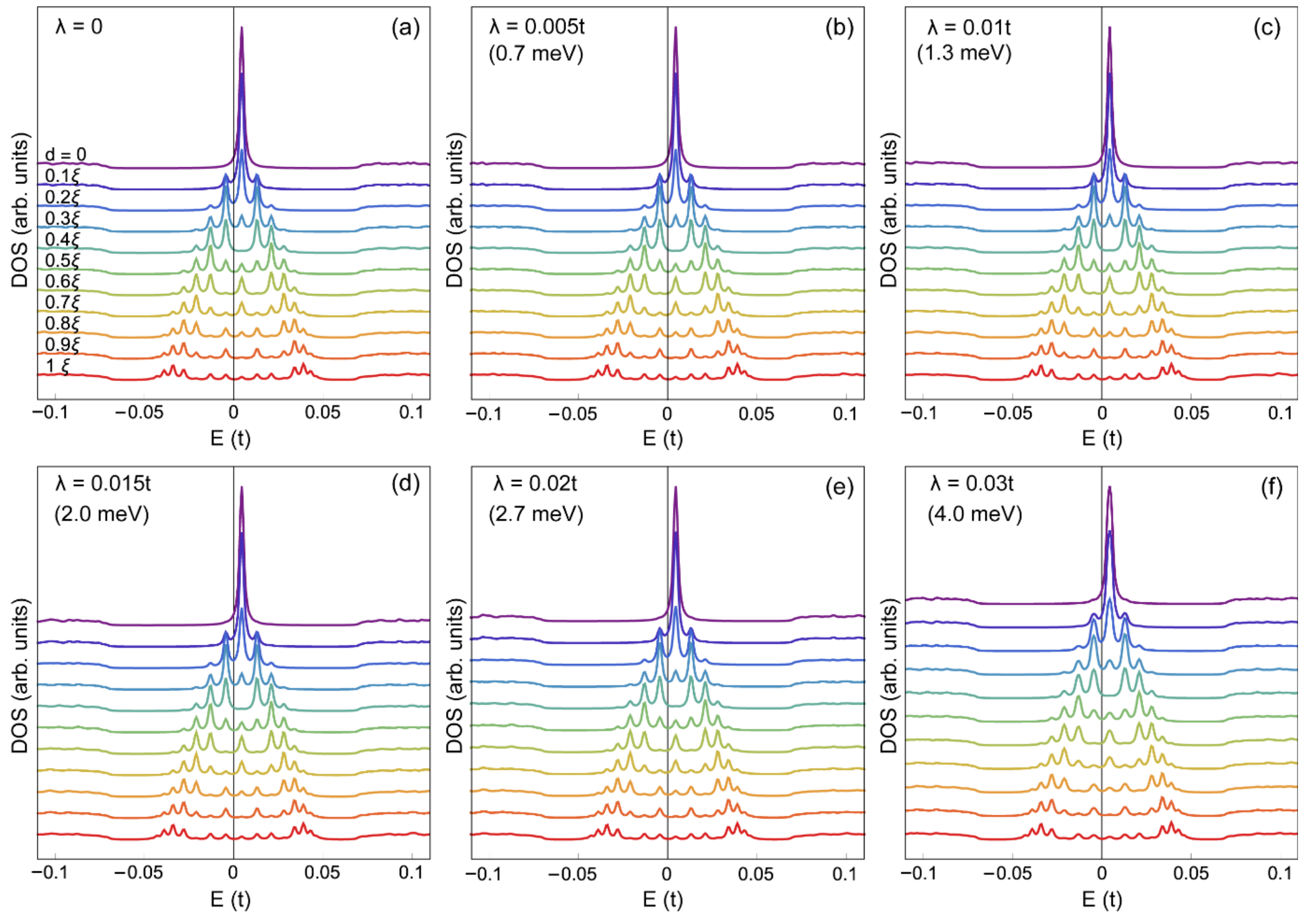

**FIG. S9.** (**a - f**) Simulated vortex states with different SOC strength ($\lambda$) under plain s-wave pairing. The curves in each panel are calculated LDOS at various distance to the vortex center (in the unit of superconducting coherence length $\xi$), as marked in (a).

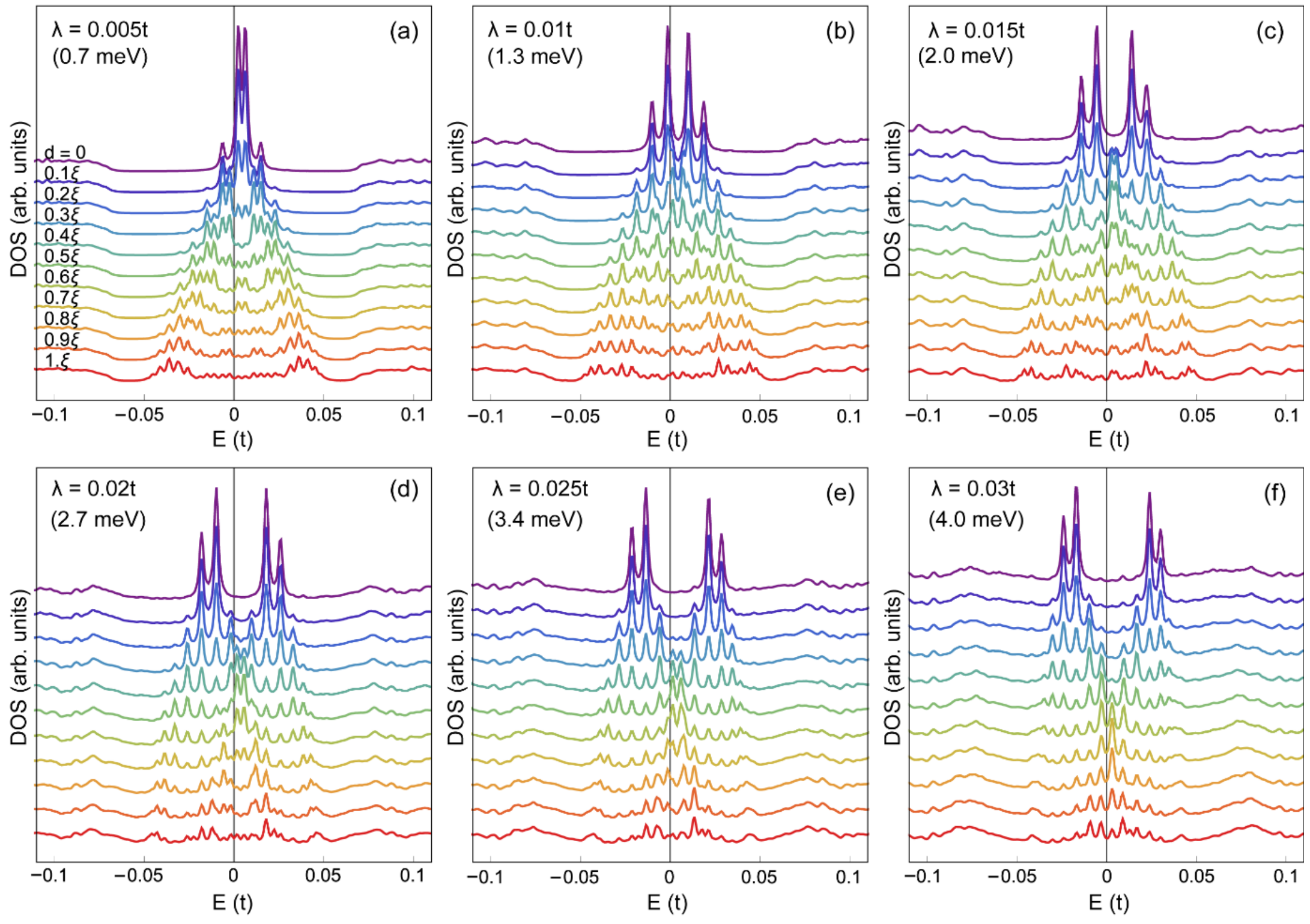

**FIG. S10.** (**a - f**) Simulated vortex states with different SOC strength ($\lambda$) under nodeless *d*-wave pairing. Curves in each panel are calculated LDOS at various distance to the vortex center, as marked in (a).

**VIII. Vortex states simulation with considering (weak) interface disorder effect and dI/dV line cut offset to the core center.**

So far, the measured energies of discrete CdGM state agree well with the s-wave pairing scenario. Nonetheless, we noticed that the *intensity* of measured CdGM states show certain asymmetric or irregular distribution with respect to core center, as seen in Fig. 2(e,f) and Fig. S6(e, f). In addition, the calculation in Fig. 4(a) shows only one peak ($E_{1/2}$) at $d$=0, however we usually observe multiple peaks near the "center" of free vortex cores (as shown in Fig. 2). In this part we demonstrate that these apparent deviations between experiment and theory could be due to randomly distributed, weak interfacial disorders in FeSe/$SrTiO_3$, and finite offset of dI/dV line cut to the core center.

For the artificial single-layer FeSe/$SrTiO_3$ system, the interface is not perfectly flat, as evidenced by that the morphology of 1ML FeSe film is always "rougher" than the thicker films (e.g., see Fig. 5(a) for comparison of 2ML and 1ML FeSe). In the free vortex regions, although we had checked the SC gap is clean (no defect induced in-gap states), there could still be weak interfacial disorders which will scatter the SC quasi-particles when vortex is generated. To study their influence to CdGM state, we consider a scalar scattering potential with a strength of U, locates at a distance $d'$ to the core center, as illustrated in Fig. S11(a). Calculated LDOS along a line cut across this scattering potential (at $d'$=0.2$\xi$) and core center are shown in Fig. S11(b~d). The calculation is based on the same two band model and *s*-wave pairing symmetry described above ($\lambda$ is set to 0 as we have shown that SOC almost has no effect on *s*-wave pairing. All other parameters are the same as that used for Fig. 4(a) ($\mu$ = -3.63$t$, $t'$ = 0.125$t$, and $t$ corresponds to ~135 meV). One can see when U=0 (Fig. S11(b)), the intensity of all the CdGM states are spatially mirror symmetric with respect to core center (d=0). After applying a small U=0.3$t$ or -0.3$t$ (~40 meV), the intensity of $E_{-1/2}$ state is clearly affected and became asymmetric with respect to core center (Fig. S11(c-d)).

Therefore, the interface disorder scattering could be the reason that the "shape" of free vortices in the dI/dV mapping are not exactly circular or fourfold symmetric (Fig. 1(d) and Fig. 2 insets). This actually makes it difficult to determine the "center" of vortex core. Moreover, the vortex core size of 1ML FeSe is rather small ($\xi$ ~2nm), therefore the measured dI/dV line cut may slightly miss the core center, as illustrated in Fig. S11(e). In Figs. S11(f~h), we show calculated line cut with various offset (0, 0.1$\xi$, 0.2$\xi$) to the core center. As the offset increases, more CdGM peaks show up in the middle of the cut. In fact this can be expected from Fig. 4(a), which show multiple peak appears at $d$ > 0.1$\xi$. Furthermore, Figs. S11(i, j, k) show the simulations with considering both line cut offset and a scalar potential on the line cut. In these cases, there are multiple CdGM states in the middle of the line cut with asymmetric intensity distribution, which reasonably reproduced our experimental observations.

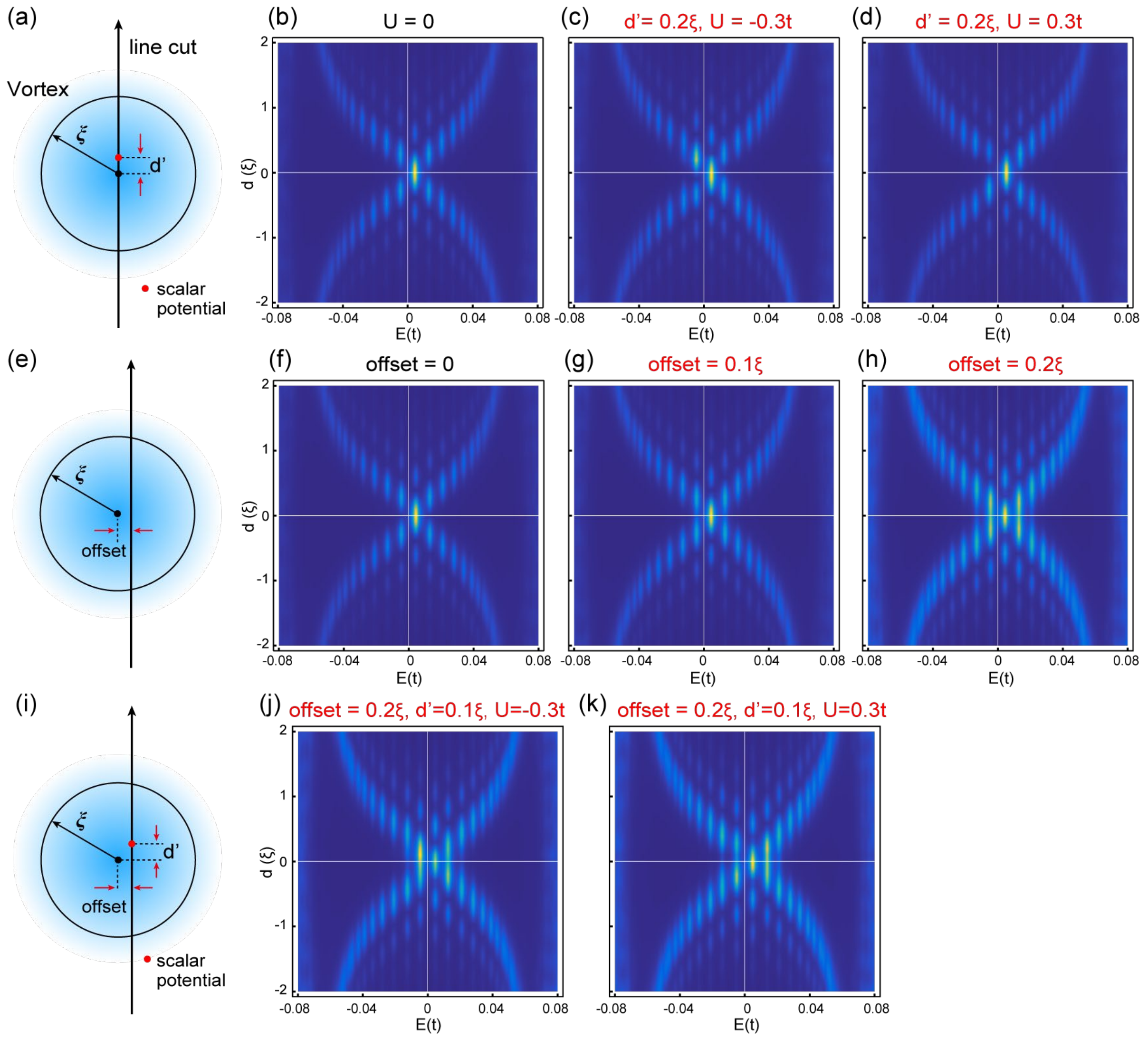

**FIG. S11. Influence of disorder scattering and line cut offset on CdGM states.** (**a**) Sketch of a vortex core (radius ~ $\xi$) and a scalar scattering potential (U) at distance of d' to the core center. (**b, c, d**) Calculated DOS along the line cut in (a) with $d' = 0.2\xi$, U = 0, -0.3$t$, and 0.3$t$, respectively. (**e**) Sketch of vortex core and an off-centered line cut. (**f, g, h**) Calculated DOS along the line cut in (e), with offset = 0, 0.1$\xi$ and 0.2$\xi$, respectively. (**i**) Sketch of vortex core with considering both disorder scattering and line cut offset. (**j, k**) Calculated DOS along the line cut in (i), with offset = 0.2$\xi$, $d' = 0.1\xi$, and U = -0.3$t$, 0.3$t$, respectively.